\documentclass[twocolumn]{aastex631}
\usepackage{amsmath}
\usepackage{hyperref}

\newcommand{\angstrom}{\text{\normalfont\AA}}

\shorttitle{Black Hole Masses in Obscured AGN}
\shortauthors{LaMassa et al.}

\graphicspath{{./}{figures/}}

\begin{document}

\title{Estimating Black Hole Masses in Obscured AGN from X-ray and Optical Emission Line Luminosities}

\correspondingauthor{Stephanie LaMassa}
\email{slamassa@stsci.edu}

\author[0000-0002-5907-3330]{Stephanie LaMassa}
\affiliation{Space Telescope Science Institute 
3700 San Martin Dr.,
Baltimore, MD, 21218, USA}

\author{Isabella Farrow}
\affiliation{Department of Physics, Yale University,
  P.O. Box 201820,
  New Haven, CT 06520-8120, USA}
\affiliation{Yale Center for Astronomy and Astrophysics,
 P.O. Box 208121, 
 New Haven, CT 06520, USA}

\author[0000-0002-0745-9792]{C. Megan Urry}
\affiliation{Department of Physics, Yale University,
  P.O. Box 201820,
  New Haven, CT 06520-8120, USA}
\affiliation{Yale Center for Astronomy and Astrophysics,
 P.O. Box 208121, 
 New Haven, CT 06520, USA}

\author[0000-0002-3683-7297]{Benny Trakhtenbrot}
\affiliation{School of Physics and Astronomy,
  Tel Aviv University, 
  Tel Aviv 69978, Israel}

\author[0000-0002-5504-8752]{Connor Auge}
\affiliation{Eureka Scientific,
  2452 Delmer Street Suite 100, 
  Oakland, CA 94602-3017, USA}

\author[0000-0002-7998-9581]{Michael J. Koss}
\affiliation{Eureka Scientific,
  2452 Delmer Street Suite 100, 
  Oakland, CA 94602-3017, USA}

\author[0000-0003-2196-3298]{Alessandro Peca}
\affiliation{Eureka Scientific,
  2452 Delmer Street Suite 100, 
  Oakland, CA 94602-3017, USA}
\affiliation{Department of Physics, Yale University,
  P.O. Box 201820,
  New Haven, CT 06520-8120, USA}

\author[0000-0002-1233-9998]{Dave Sanders}
\affiliation{Institute for Astronomy,
  University of Hawaii, 
  2680 Woodlawn Drive, Honolulu, HI, 96822, USA}

\author[0000-0003-2971-1722]{Tracey Jane Turner}
\affiliation{Eureka Scientific,
  2452 Delmer Street Suite 100, 
  Oakland, CA 94602-3017, USA}

\begin{abstract}
 We test a novel method for estimating black hole masses ($M_{\rm BH}$) in obscured active galactic nuclei (AGN) that uses proxies to measure the full-width half maximum of broad H$\alpha$ (FWHM$_{\rm bH\alpha}$) and the accretion disk luminosity at 5100 $\angstrom$ ($\lambda L_{\rm 5100\angstrom}$). Using a published correlation, we estimate FWHM$_{\rm bH\alpha}$ from the narrow optical emission line ratio $L_{\rm [O\,III]}/L_{\rm nH\beta}$. Using a sample of 99 local obscured AGN from the Swift-BAT AGN Spectroscopic Survey, we assess the agreement between estimating $\lambda L_{\rm 5100\angstrom}$ from the intrinsic 2-10 keV X-ray luminosity and from narrow optical emission lines. We find a mean offset of $0.32 \pm 0.68$ dex between these methods, which propagates to a factor of $\sim$2 uncertainty when estimating $M_{\rm BH}$ using a virial mass formula where $L_{\rm [O\,III]}/L_{\rm nH\beta}$ serves as a proxy of  FWHM$_{\rm bH\alpha}$ ($M_{\rm BH,[O\,III]/nH\beta}$). We compare $M_{\rm BH,[O\,III]/nH\beta}$ with virial $M_{\rm BH}$ measurements from broad Paschen emission lines.  For the 14 (12) BASS AGN with broad Pa$\alpha$ (Pa$\beta$) detections, we find $M_{\rm BH,[O\,III]/nH\beta}$ to be systematically higher than $M_{\rm BH,Pa\alpha}$ ($M_{\rm BH,Pa\beta}$) by a factor of 0.39 $\pm$ 0.44 dex (0.48 $\pm$ 0.51 dex). Since these offsets are within the scatter, more data are needed to assess whether $M_{\rm BH,[O\,III]/nH\beta}$ is biased high. For 151 BASS AGN with measured stellar velocity dispersions ($\sigma_{\rm *}$), we find that the $\sigma_{\rm *}$-derived $M_{\rm BH}$ agrees with $M_{\rm BH,[O\,III]/nH\beta}$ to within 0.08 dex, albeit with wide scatter (0.74 dex). The method tested here can provide estimates of $M_{\rm BH}$ in thousands of obscured AGN in spectroscopic surveys when other diagnostics are not available, though with an uncertainty of $\sim$3-5.
\end{abstract}


\section{Introduction}

The mass of a black hole ($M_{\rm BH}$) is one if its fundamental properties \citep{misner}. A primary mechanism of black hole growth is via accretion of nearby matter. Steady accretion onto supermassive black holes (SMBH; $M_{\rm BH} \geq 10^{6}$ M$_{\rm \odot}$) in the centers of galaxies produce Active Galactic Nuclei (AGN) which are the most luminous, persistent objects in the Universe \citep{osterbrock1993}. Studying AGN thus provides us a way to constrain the growth history of massive black holes over cosmic time, as long as the mass of the black hole can be measured.

In nearby non-active galaxies where the central region of the galaxy can be resolved, the SMBH mass can be estimated by measuring the kinematics of the stars and/or gas moving around the black hole. Indeed, decades of mapping out the motions of stars in the center of our own Milky Way galaxy provided the strongest evidence yet of a quiescent black hole, Sagittarius A*, of mass $4 \times 10^{6}$ M$_{\rm \odot}$ \citep{ghez1998,ghez2008,genzel2000,genzel2010}. Applying kinematic analysis to neighboring galaxies provides dynamical measurements of SMBH masses, along with the discovery of a correlation between properties of the host galaxy (velocity stellar dispersion, bulge luminosity, total stellar mass) and the SMBH mass \citep{kormendy1995,magorrian1998,ferrarese2000,ferrarese2001,gebhardt2000, gultekin2009, kormendy2013, reines2015}.

These methods are inapplicable to unobscured AGN as emission from the accretion disk dominates over that of the host galaxy, precluding the ability to resolve and measure the motion of stars or gas within the sphere of influence of the black hole. The motion of gas near the black hole ionized by the accretion disk can be used to derive the $M_{\rm BH}$ as long as the distance to these gas clouds can be measured. From reverberation mapping, we can measure how the broad emission line profiles from ionized gas in the Broad Line Region (BLR) respond to accretion driven luminosity fluctuations \citep[see][for a review]{peterson2014}. These reverberation mapping campaigns are necessarily time intensive, spanning months to years, and though have accurately measured AGN black hole masses for hundreds of AGN \citep[e.g.][]{shen2015,shen2019,shen2024,grier2017,grier2019,homayouni2020}, this method does not easily scale to tens of thousands of AGN.

But these campaigns have been critical for establishing that the accretion disk luminosity serves as a proxy of the distance to the black hole \citep[e.g.,][]{bentz2006,bentz2009,kaspi2005}. The gas orbital speed can be estimated by the width of emission lines in the BLR, allowing the black hole mass to be measured from single-epoch spectra ($M_{\rm BH} = f \frac{R v^2}{G}$; $R$ - distance between black hole and ionized gas, $v$ - gas orbital speed, $f$ - scale factor that depends on the kinematics and geometry of the broad line region, $G$ - gravitational constant). Different single-epoch spectra virial mass relationships have been calibrated for emission lines from the ultraviolet \citep[\ion{C}{4} 1549$\angstrom$, \ion{Mg}{2} 2800$\angstrom$;][]{vestergaard2006,shen2011,trakhtenbrot2012}, through the optical \citep[H$\beta$ 4861$\angstrom$, H$\alpha$ 6563$\angstrom$;][]{vestergaard2006,greene2010}, to the near-infrared \citep[Pa$\beta$ 12822$\angstrom$, Pa$\alpha$ 18751$\angstrom$;][]{kim2010}. Each mass scaling relationship has an uncertainty of 0.18 - 0.6 dex, and the agreement between $M_{\rm BH}$ values calibrated from different emission lines can vary by a factor of 0.12 - 0.4 dex \citep{shen2012,trakhtenbrot2012}.

These methods of estimating black hole masses only work when we have a direct view of the accretion disk and broad line region. Most AGN are obscured \citep{ramosalmeida2017,hickox2018} such that this central region is blocked by large amounts of dust and gas. One of the best ways to measure SMBH masses in obscured AGN is to measure accretion disk Keplarian velocities from water maser emission at $\lambda = 1.35$ cm \citep{moran1999}, but only a small percentage of AGN are in favorable conditions and nearby for water maser emission to be detected \citep{zhu2011}. A common way to estimate black hole masses in obscured AGN is to leverage the correlation between the dynamical mass of SMBHs and velocity dispersion of stars in the galaxy ($\sigma_{\rm *}$), which has an intrinsic scatter of $\sim$0.3 dex \citep[see][for a review]{kormendy2013}, though the spread can be as high as $\sim$0.5 dex \citep{marsden2020}. Accurate measurements of the stellar velocity dispersion requires high quality spectra which is prohibitively difficult to obtain for large samples of AGN beyond the very local Universe \citep[$z > 0.1$;][]{koss2022b}.

A promising technique to estimate black hole masses in obscured AGN was presented by \citet[][hereafter BM19]{baron}. Using a sample of 1941 unobscured (Type 1) AGN at $z < 0.3$ from Data Release 7 of the Sloan Digital Sky Survey \citep[SDSS;][]{york2000,sdssdr7}, they found that the AGN ionization field hardness (traced by the ratio of the narrow [\ion{O}{3}]$\lambda$ 5007 to narrow H$\beta$ line, $L_{\rm [O\,III]}/L_{\rm nH\beta}$) in the Narrow Line Region (NLR) correlates with BLR kinematics (parameterized by the Full-Width Half Maximum, FWHM, of broad H$\alpha$, FWHM$_{\rm bH\alpha}$), though it is unclear what physically links the BLR kinematics to the NLR ionization to give rise to this correlation.  Since the key ingredients for measuring M$_{\rm BH}$ in virial mass formulas are the BLR kinematics and accretion disk luminosity, this empirical correlation found by BM19 implies that $L_{\rm [O\,III]}/L_{\rm nH\beta}$ can be used as a proxy of FWHM$_{\rm bH\alpha}$ to calculate $M_{\rm BH}$ (hereafter $M_{\rm BH,[O\,III]/nH\beta}$). The other key parameter in virial black hole mass formulas, the accretion disk luminosity, is invisible in obscured AGN.  BM19 suggest that the accretion disk luminosity at 5100$\angstrom$ ($\lambda L_{\rm 5100\angstrom}$) can be estimated from either narrow optical emission lines \citep{netzer2009} or the intrinsic 2-10 keV luminosity \citep{maiolino2007} in order to calculate $M_{\rm BH,[O\,III]/nH\beta}$.

BM19 applied their technique to calculate black hole masses for 10,000 obscured (Type 2) AGN from SDSS where they estimated $\lambda L_{\rm 5100\angstrom}$ from narrow optical emission lines and FWHM$_{\rm bH\alpha}$ from $L_{\rm [O\,III]}/L_{\rm nH\beta}$. When comparing these black hole masses with the host galaxy stellar velocity dispersion, they found a scatter of $\sim$0.45 dex, comparable to quoted uncertainties in $M_{\rm BH}$ - $\sigma_{\rm *}$ relations \citep{kormendy2013,marsden2020}, leading them to conclude that $M_{\rm BH,[O\,III]/nH\beta}$ is a reliable way to estimate black hole mass. Some studies have since used the BM19 method to calculate $M_{\rm BH,[O\,III]/nH\beta}$ for Type 2 AGN \citep{rey2021,ferre-mateu2021,vietri2022,siudek2023} and a larger number of studies have compared theoretical predictions of $M_{\rm BH}$/host galaxy correlations with the BM19 $M_{\rm BH,[O\,III]/nH\beta}$ sample \citep{shankar2020,volenteri2020,habouzit2021,dubois2021,treibitsch2021,trinca2022,sassano2023,beckmann2023}. Though the BM19 formulation to estimate $M_{\rm BH}$ is being adopted, this technique has only been tested on indirect methods of measuring $M_{\rm BH}$ (i.e., correlating $M_{\rm BH}$ with $\sigma_{\rm *}$). \citet{baron} caution that their technique can not be reliably tested on Type 1 AGN due to the uncertainty in decomposing the narrow H$\beta$ emission line from the broad component. They instead point out that a direct test would be to identify a sample of AGN with no broad H$\beta$ emission but with broad Pa$\alpha$ or Pa$\beta$ lines detected from which the virial black hole mass can be calculated \citep{kim2010,landt2011a,landt2011b,landt2013}. Since the Paschen series lie in the near-infrared, these lines are less affected by dust obscuration that extinguishes the optical emission, so it is possible for some AGN to show both narrow Balmer lines and broad Paschen lines.

We perform this test using local ($z \leq 0.3$) obscured AGN detected from the the ultra-hard X-ray (14 - 195 keV) Swift-BAT sample, which has extensive optical and infrared spectroscopic data available from the BAT AGN Spectroscopic Survey \citep[BASS;][]{koss2017,koss2022a,oh2022,riccif2017,lamperti2017,denbrok2022}. Using this dataset, we compare estimates of $\lambda L_{\rm 5100\angstrom}$ derived from the intrinsix X-ray luminosity between 2-10 keV \citep[$L_{\rm 2-10 keV}$;][]{maiolino2007,lusso2010} and from the optical emission line luminosities \citep{netzer2009}. We note that the empirical relationships we are testing to estimate FWHM$_{\rm bH\alpha}$ and $\lambda L_{\rm 5100\angstrom}$ were calibrated on Type 1 AGN and we are applying these relationships to obscured AGN which may not be appropriate if there are systematic differences in these quantities between unobscured and obscured AGN. However, our intention is to investigate how well these empirical relationships perform when estimating $M_{\rm BH}$ in obscured AGN: any systematic differences will contribute to the observed spread and quoted uncertainty in black hole mass measurements. We test whether a correlation exists between $L_{\rm [O\,III]}/L_{\rm nH\beta}$ and the BLR kinematics traced by the Pa$\alpha$ and Pa$\beta$ FWHM values. We then calculate the black hole masses for AGN using the single-epoch spectrum virial mass equations for Pa$\alpha$ and Pa$\beta$ and compare those with $M_{\rm BH,[O\,III]/nH\beta}$. Finally, for a larger sample of BASS AGN, we compare $M_{\rm BH,[O\,III]/nH\beta}$ with black hole masses derived from $\sigma_{\rm *}$.  For reference, we use a flat cosmology where H$_{\rm 0}$ = 67.7 km/s/Mpc and $\Omega_{m}$ = 0.307 \citep{planck2015}.

\section{Sample Selection}
We select AGN from the BASS sample \citep{koss2017,koss2022a} for this analysis due to the homegeneity, multi-wavelength completeness, and quality of this dataset. The Burst Alert Telescope \citep[BAT;][]{barthelmy2005} on-board the Neil Gehrels Swift Observatory \citep{gehrels2004} is sensitive to ultra-hard X-rays (14-195 keV) and has detected hundreds of AGN. The BASS survey provides extensive multi-wavelength follow-up of these ultra-hard X-ray selected AGN, ranging from near-infrared \citep[NIR;][]{riccif2017,lamperti2017,denbrok2022} and optical \citep[e.g.,][]{oh2022,koss2022b} spectroscopy to detailed X-ray analysis using softer X-ray data (0.5 - 10 keV) from {\it Chandra}, {\it XMM-Newton}, and Swift-XRT \citep{riccic2017}.

To begin, we identify BASS AGN that have [\ion{O}{3}]$\lambda$ 5007, narrow H$\beta$ (nH$\beta$), and intrinsic 2-10 keV X-ray fluxes measured, as well as no broad component to H$\beta$ detected \citep{oh2022,riccic2017}. Details about fitting the optical emission lines are provided in \citet{oh2022}, but in short: the spectra were initially fitted with narrow emission lines and when the spectra were not well described by these fits, broad Gaussian components with a FWHM above 1000 km/s were added. \citet{oh2022} report fluxes for emission lines if the Gaussian amplitude over noise ratio exceeded 3.\footnote{Amplitude over noise ratios are a typical metric used to claim an emission line as significant. The amplitude is the height of the Gaussian function and noise is determined from the root mean square between the data and the model in the continuum on either side of the emission line.} They provide a flag in their published tables to indicate whether a broad Balmer line is used in the spectral fit. We filtered their table to identify the AGN where they flagged that no broad component was used to fit H$\beta$.

BM19 demonstrate that the correlation they found between $L_{\rm [O\,III]}/L_{\rm nH\beta}$ and FWHM$_{\rm bH\alpha}$ holds for $\log$($L_{\rm [O\, III]}/L_{\rm nH\beta}$) $>$ 0.55 dex which is the cut-off at which AGN ionization dominates over star-formation in powering the [\ion{O}{3}] line flux \citep{bpt,kewley2001}. Their sample has an upper limit of  $\log$($L_{\rm [O\, III]}/L_{\rm nH\beta}$) $<$ 1.05 dex. We apply both cuts on the ionization hardness to the BASS dataset, giving us a parent sample of 172 AGN (Table \ref{sample_numbers}).

For our first test, we compare the agreement between two methods of estimating the accretion disk luminosity at 5100 $\angstrom$ (i.e., $\lambda L_{\rm 5100 \angstrom}$). In obscured AGN, the accretion disk is blocked from view and $\lambda L_{\rm 5100 \angstrom}$ can therefore not be directly measured. However, there are methods to {\it estimate} $\lambda L_{\rm 5100 \angstrom}$ based on the intrinsic X-ray luminosity \citep{kaspi2005,maiolino2007,lusso2010} and optical narrow emission line fluxes \citep{netzer2009}. This latter method requires a reddening correction calculated from the ratio of the narrow H$\alpha$ (nH$\alpha$) and nH$\beta$ lines. We therefore identify a subsample where nH$\alpha$ is detected without a broad H$\alpha$ component. This selection provides 99 AGN for quantifying the consistency between $\lambda L_{\rm 5100 \angstrom}$ proxies.

We assess the consistency between black hole masses calculated using the BM19 method with measurements of the virial mass calculated from broad Pa$\alpha$ and Pa$\beta$ emission lines.  Using data from \citet{lamperti2017,riccif2022,denbrok2022}, we identify AGN where the quality of the fit to the Paschen lines was deemed very good (\textsc{flag} == 1) or acceptable (\textsc{flag} == 2).\footnote{Only the catalogs of \citet{lamperti2017} and \citet{riccif2022} provide this spectral fitting information. We include all Pa$\alpha$ or Pa$\beta$ measurements from the \citet{denbrok2022} catalog under the assumption that all of these published values are reliable.}  Of the 172 AGN in the BASS parent sample, we find 14 and 12 AGN with broad Pa$\alpha$ and broad Pa$\beta$ detected, respectively; 8 AGN have both broad Pa$\alpha$ and broad Pa$\beta$ detected.Finally, we use the measured stellar velocity dispersions of 151 BASS AGN \citep{koss2022b} to indirectly estimate $M_{\rm BH}$ from the $M_{\rm BH} - \sigma_{\rm *}$ relation, and compare those estimtes with $M_{\rm BH}$ measured from the BM19 method.

The optical spectra were collected from telescopes with aperture slit widths between 1$^{\prime\prime}$-3$^{\prime\prime}$ \citep{koss2022c}. The projected size of the optical spectroscopic apertures cover $>$600 pc of the AGN host galaxy for over 78\% of the BASS AGN. This scale is consistent with the physical scales probed by the SDSS aperture in the BM19 AGN sample (0.6 kpc - 12 kpc), making the BASS sample an appropriate one to test the efficacy of a relationship calibrated on SDSS AGN. We note that the infrared spectroscopic apertures cover smaller scales \citep[projected sizes of 0.09 - 1.6 kpc;][]{lamperti2017,riccif2022,denbrok2022}, but here we are probing emission from the compact broad line region where smaller aperture are appropriate to minimize host galaxy dilution.

\begin{deluxetable*}{llllll}
\tablecaption{\label{sample_numbers}Number of AGN in Each Subsample}
\tablehead{\colhead{Sample} & \colhead{$M_{\rm BH,[O\,III]/nH\beta}$} & \colhead{$\lambda L_{\rm 5100 \angstrom}$} & \colhead{Broad Pa$\alpha$} & \colhead{Broad Pa$\beta$} & \colhead{$\sigma_{\rm *}$}
}
\startdata
$M_{\rm BH,[O\,III]/nH\beta}$\tablenotemark{a} & 172 & 99 & 14 & 12 & 151 \\
$\lambda L_{\rm 5100 \angstrom}$\tablenotemark{b} & 99 & 99 & 2 & 2 & 94\\
Broad Pa$\alpha$\tablenotemark{c} & 14 & 2 & 14 & 8 & 12\\
Broad Pa$\beta$\tablenotemark{d} &  12 & 2 & 8 & 12 & 10\\
$\sigma_{\rm *}$\tablenotemark{e} & 151 & 94 & 12 & 10 & 151\\
\enddata
\tablenotetext{a}{Parent sample of BASS AGN with [\ion{O}{3}] detected, narrow H$\beta$ detected, intrinsic 2-10 keV flux measured, no broad H$\beta$ component detected, and within the ionization hardness range of 0.55 dex $<$ $\log$($L_{\rm [O\,III]}/L_{\rm nH\beta}$) $<$ 1.05.}
\tablenotetext{b}{Subset of parent sample with narrow H$\alpha$ detected and no broad H$\alpha$ component detected, used to compare proxies for estimating $\lambda L_{\rm 5100 \angstrom}$.}
\tablenotetext{c}{Subset of parent sample with broad Pa$\alpha$ detected.}
\tablenotetext{d}{Subset of parent sample with broad Pa$\beta$ detected.}
\tablenotetext{e}{Subset of parent sample with measured stellar velocity dispersions.}
\end{deluxetable*}

In Table \ref{bass_id_table}, we identify the BASS AGN used for this analysis and the sample(s) in which they belong. We list the emission line fluxes, intrinsic 2-10 keV X-ray fluxes, and infrared fluxes and FWHM values in Table \ref{bass_flux_table}. 

\begin{deluxetable*}{lllllcccc}
\tablecaption{\label{bass_id_table}BASS Analysis Sample\tablenotemark{1}}
\tablehead{\colhead{ID} & \colhead{Swift-BAT Name} & \colhead{RA} & \colhead{Dec} & \colhead{z} & \colhead{$\lambda L_{\rm 5100 \angstrom}$} & \colhead{Pa$\alpha$} & \colhead{Pa$\beta$} & \colhead{$\sigma_{\rm *}$}\\
& & & & & \colhead{sample} & \colhead{sample} & \colhead{sample} & \colhead{sample} }
\startdata
63 & SWIFTJ0114.4-5522 & 18.6039 & -55.397 & 0.0124 & N & N & Y & Y \\
72 & SWIFTJ0123.8-3504 & 20.9765 & -35.065 & 0.019 & N & Y & Y & Y \\
226 & SWIFTJ0433.0+0521 & 68.2962 & 5.354 & 0.0334 & N & Y & Y & N \\
404 & SWIFTJ0804.2+0507 & 121.024 & 5.114 & 0.013 & Y & Y & N & Y \\
511 & SWIFTJ1042.4+0046 & 160.535 & 0.702 & 0.095 & N & Y & Y & Y \\
586 & SWIFTJ1204.5+2019 & 181.124 & 20.316 & 0.0229 & N & Y & N & Y \\
682 & SWIFTJ1338.2+0433 & 204.566 & 4.543 & 0.023 & N & N & Y & Y \\
698 & SWIFTJ1353.7-1122 & 208.368 & -11.385 & 0.0687 & N & Y & N & Y \\
700 & SWIFTJ1354.5+1326 & 208.621 & 13.466 & 0.063 & Y & Y & N & Y \\
738 & SWIFTJ1441.4+5341 & 220.159 & 53.504 & 0.0377 & N & Y & N & Y \\
757 & SWIFTJ1508.8-0013 & 227.225 & -0.197 & 0.0545 & N & Y & Y & Y \\
971 & SWIFTJ1824.2+1845 & 276.045 & 18.769 & 0.0663 & N & Y & N & Y \\
1027 & SWIFTJ1913.3-5010 & 288.311 & -50.183 & 0.062 & Y & N & Y & Y \\
1060 & SWIFTJ2001.0-1811 & 300.232 & -18.174 & 0.037 & N & Y & Y & Y \\
1090 & SWIFTJ2044.2-1045 & 311.041 & -10.724 & 0.0347 & N & Y & Y & N \\
1138 & SWIFTJ2204.7+0337 & 331.08 & 3.564 & 0.0611 & Y & N & Y & Y \\
1157 & SWIFTJ2235.9-2602 & 338.943 & -26.05 & 0.0049 & N & Y & Y & Y \\
1161 & SWIFTJ2236.7-1233 & 339.194 & -12.545 & 0.024 & N & Y & Y & Y \\
\enddata
\tablenotetext{1}{The BASS AGN sample is presented in \citet{koss2017,koss2022a}. We use optical emission line fluxes from \citet{oh2022}, near-infrared emission line fluxes from \citet{lamperti2017,riccif2022,denbrok2022} and X-ray fluxes from \citet{riccic2017}. We show here a subset of this table for illustrative purposes. The catalog is available in its entirety on-line.}
\end{deluxetable*}

\begin{deluxetable*}{lllllllll}
\tablecaption{\label{bass_flux_table}BASS AGN Fluxes and Paschen FWHM values\tablenotemark{1}}
\tablehead{\colhead{ID} & \colhead{$F_{\rm nH\beta}$} & \colhead{$F_{\rm [O\,III]}$} & \colhead{$F_{\rm H\alpha}$} & \colhead{$F_{\rm Pa\alpha}$} & \colhead{FWHM$_{\rm Pa\alpha}$} & \colhead{$F_{\rm Pa\beta}$} & \colhead{FWHM$_{\rm Pa\beta}$} & \colhead{$F_{\rm 2-10keV,intrinsic}$} \\
  & \colhead{10$^{-14}$ erg s$^{-1}$}  & \colhead{10$^{-14}$ erg s$^{-1}$}  & \colhead{10$^{-14}$ erg s$^{-1}$} & \colhead{10$^{-14}$ erg s$^{-1}$} & \colhead{km s$^{-1}$} & \colhead{10$^{-14}$ erg s$^{-1}$} & \colhead{km s$^{-1}$} & \colhead{10$^{-12}$ erg s$^{-1}$}}
\startdata
63 & 0.53 & 2.66 & \nodata & \nodata & \nodata & 1.58 & 1430 & 4.50 \\
72 & 0.78 & 8.41 & \nodata & 17.4 & 5220 & 6.59 & 6670 & 22.4 \\
226 & 9.47 & 49.0 & \nodata & 33.6 & 2606 & 23.9 & 2842 & 38.2 \\
404 & 8.02 & 64.4 & 29.0 & 1.90 & 3156 & \nodata & \nodata & 32.9 \\
511 & 0.28 & 1.60 & \nodata & 1.53 & 2520 & 0.50 & 1320 & 3.10 \\
586 & 1.35 & 11.5 & \nodata & 2.39 & 3970 & \nodata & \nodata & 7.00 \\
682 & 1.60 & 8.56 & \nodata & \nodata & \nodata & 2.89 & 6189 & 10.8 \\
698 & 0.15 & 1.16 & \nodata & 1.19 & 6449 & \nodata & \nodata & 6.00 \\
700 & 0.41 & 4.23 & 1.36 & 0.53 & 6159 & \nodata & \nodata & 4.10 \\
738 & 9.39 & 50.9 & \nodata & 2.30 & 2173 & \nodata & \nodata & 5.50 \\
757 & 1.41 & 14.5 & \nodata & 4.10 & 6783 & 9.09 & 8229 & 8.10 \\
971 & 0.99 & 8.47 & \nodata & 5.49 & 9446 & \nodata & \nodata & 4.10 \\
1027 & 0.06 & 0.28 & 0.32 & \nodata & \nodata & 0.88 & 2720 & 8.40 \\
1060 & 0.21 & 1.32 & \nodata & 17.7 & 4412 & 7.70 & 4362 & 6.30 \\
1090 & 8.14 & 37.0 & \nodata & 61.4 & 3133 & 43.8 & 2918 & 44.1 \\
1138 & 2.16 & 17.4 & 8.64 & \nodata & \nodata & 0.84 & 3220 & 7.20 \\
1157 & 0.56 & 4.30 & \nodata & 4.30 & 1570 & 2.84 & 1420 & 41.8 \\
1161 & 6.15 & 46.4 & \nodata & 5.53 & 2571 & 4.43 & 4325 & 11.7 \\
\enddata
\tablenotetext{1}{The optical emission line measurements are from \citet{oh2022}, the near-infrared line measurements are from \citet{lamperti2017,riccif2022,denbrok2022}, and the X-ray flux is from \citet{riccic2017}. We show here a subset of this table for illustrative purposes. The catalog is available in its entirety on-line.}
\tablenotetext{2}{The [\ion{O}{3}] flux reported in \citet{oh2022} has a reddening correction applied. In our analysis, we back out this correction using their reported correction factor so that we are using the observed [\ion{O}{3}] flux.}
\end{deluxetable*}

\section{Analysis}

To calculate the black hole masses in obscured AGN, we use Equation 5 from BM19 which was derived using the radius-luminosity relation reported in \citet{bentz2013} and the correlation between FWHM$_{\rm bH\alpha}$ and FWHM$_{\rm bH\beta}$ derived in \citet{greene2005}:
\begin{equation}\label{o3_hb_bhmass_formula}
  \begin{split}
    {\rm \log} \left(\frac{M_{\rm BH}}{M_{\odot}}\right) = \rm{\log}\, \epsilon + 6.90\, +\\
    \, 0.54 \times \, \rm{\log} \left(\frac{\lambda L_{\rm 5100 \angstrom}}{10^{44}\, {\rm erg \, s^{-1}}} \right) \\
    + 2.06 \, \times \, \rm{\log} \left(\frac{{\rm FWHM_{\rm bH\alpha}}}{10^{3} \, {\rm km \, s^{-1}}}\right)
    \end{split}
\end{equation}\\

Here, we assume the scaling factor ($\epsilon$) in their Equation 5 is unity. While it is challenging to determine the geometric scale factor for individual AGN due to unknowns in the BLR geometry, inclination, and kinematics, the average value for this virial factor is approximately unity when using the FWHM of broad emission lines to calculate $M_{\rm BH}$ \citep[e.g., see][]{woo2015}. Published papers from the BASS survey adopt a geometric scaling factor of unity when calculating $M_{\rm BH}$ where the virial motion of the gas is measured from the FWHM of the broad line \citep[note that $\epsilon=1$ corresponds to a virial factor of 5.5 when using the line velocity dispersion as a proxy of gas orbital speed;][]{mejia-restrepo2022, denbrok2022} and we adopt this convention in the current analysis for consistency, especially when comparing our $M_{\rm BH,[O\,III]/nH\beta}$ values with those derived from NIR Paschen virial mass equations. We note that this scale factor is calculated for Type 1 AGN where the inclination of the BLR is putatively aligned face-on and we are assuming this factor is a reasonable starting point to estimate $M_{\rm BH}$ in edge-on AGN: such uncertainties in the geometric scale factor will contribute to uncertainties in the final $M_{\rm BH,[O\,III]/nH\beta}$ values.

To solve Equation \ref{o3_hb_bhmass_formula}, we need both an estimate of the accretion disk luminosity $\lambda L_{\rm 5100 \angstrom}$ and the full-width half maximum of the broad H$\alpha$ line.  FWHM$_{\rm bH\alpha}$ can be derived from the $L_{\rm [O\,III]}/L_{\rm nH\beta}$ ratio using Equation 1 from BM19 which we recast as:
\begin{equation}
  \begin{split}\label{hafwhm_eq}
    \rm{\log} \left(\frac{FWHM_{\rm bH\alpha}}{ {\rm 10^3 \, km \, s^{-1}}} \right) = (1.72 \, \pm \, 0.21) \, \times \\
    \rm{\log} \left(\frac{L_{\rm [OIII]}}{L_{\rm nH\beta}}\right) - (0.62 \pm 0.19)
    \end{split}
\end{equation}\\

\subsection{Estimating Accretion Disk Luminosity at 5100 $\angstrom$ ($\lambda L_{\rm 5100\angstrom}$)}

The ultraviolet to optical emitting accretion disk is hidden in obscured AGN meaning that $\lambda L_{\rm 5100 \angstrom}$ can not be measured directly from the optical spectrum. There are proxies for estimating $\lambda L_{\rm 5100 \angstrom}$ that are derived by relations observed in unobscured AGN. One proxy comes from the optical narrow emission lines (herein dubbed $\lambda L_{\rm 5100 \angstrom,nel}$) and another comes from the intrinsic 2-10 keV luminosity (herafter $\lambda L_{\rm 5100 \angstrom,X-ray}$). Both methods have advantages and drawbacks in their applicability to AGN detected in surveys.

The optical method requires a reddening correction to the narrow line to be calculated via the Balmer decrement (nH$\alpha$/nH$\beta$). This method is applicable to thousands of AGN from ground-based optical surveys at $z < 0.5$ (beyond which H$\alpha$ is redshifted into the infrared) and can be used for infrared spectroscopic studies of higher redshift AGN up to $z < 1$, after which the $L_{\rm [O\,III]}/L_{\rm nH\beta}$ ratio for star-forming galaxies increases \citep{kewley2013}.

The X-ray method can be a powerful technique to use for tens of thousands of AGN detected in X-ray surveys and can in principle be used at any redshift. However, this method requires that the intrinsic X-ray luminosity be known via high fidelity spectral modeling which is hard to accurately measure for AGN with low X-ray flux, including highly obscured AGN and those at high redshift, due to a combination of generally low spectral quality and the number of model parameters needed to constrain the obscuring column density. Due to the proximity of the BASS AGN and the significant time investment in observing these sources with X-ray facilities, the spectra of many of these AGN were of high quality to permit sophisticated spectral modeling \citep[see][for details]{riccic2017}.

Here, we test the agreement between these two methods of estimating $\lambda L_{\rm 5100\angstrom}$ in obscured AGN.

\subsubsection{$\lambda L_{\rm 5100\angstrom,nel}$ : Estimating $\lambda L_{\rm 5100,\angstrom}$ from Optical Narrow Emission Lines}

First we derive a formula for $\lambda L_{\rm 5100\angstrom,nel}$ using previous published relationships that link the 5100 $\angstrom$ luminosity to the bolometric luminosity, and the bolometric luminosity to optical narrow emission lines. Using the relationship found between $\lambda L_{\rm 5100 \angstrom}$ and the isotropic AGN luminosity ($L_{\rm iso}$), we have from \citet{runnoe2012}:

\begin{equation}\label{l5100_opt}
  {\rm\log} \left(\frac{\lambda L_{\rm 5100,optical}}{10^{44} \, {\rm erg\, s^{-1}}}\right) = 1.0965 \times {\rm \log} \left(\frac{L_{\rm bol}}{10^{44} \, {\rm erg\, s^{-1}}}\right) - 0.98.
\end{equation}

This formula assumes  $L_{\rm bol} \approx 0.75 L_{\rm iso}$ to correct for anisotropy in the viewing angle of the accretion disk \citep[see][]{runnoe2012}.

\citet{netzer2009} demonstrated that $L_{\rm bol}$ can be estimated using the luminosity of the reddening-corrected narrow H$\beta$ line ($L_{\rm nH\beta,corr}$) and the ratio of the observed [\ion{O}{3}] luminosity to observed nH$\beta$ luminosity:
\begin{equation}\label{lbol}
  \begin{split}
    {\rm \log} \left(\frac{L_{\rm bol}}{10^{44} \, {\rm erg\,s^{-1}}}\right) = {\rm \log} \left(\frac{L_{\rm nH\beta,corr}}{10^{40} \, {\rm erg \, s^{-1}}}\right) \, - \, 0.25 + \\
    {\rm max}\left[0,0.31\left({\rm \log}\frac{L_{\rm [OIII]}}{L_{\rm nH\beta}} - 0.6\right)\right].
  \end{split}
\end{equation}

We emphasize that this equation assumes a Galactic reddening relation \citep{cardelli1989} rather than the $\lambda^{-0.7}$ extinction law from \citep{charlot2000} that was considered in \citet{netzer2009}. We note that the equation in BM19 uses the $\lambda^{-0.7}$ extinction law, but then a Galactic extinction curve to derive the reddening correction to the narrow H$\beta$ line. Here, we assume the Galactic extinction law throughout for internal consistency and for consistency with previous optical analysis of obscured AGN \citep[e.g.,][]{bassani1999,lamassa2009,lamassa2010, oh2022}.

Assuming Case B recombination in the AGN narrow line region \citep{osterbrock2006} and the extinction curve of \citet{cardelli1989}, we calculate the reddening corrected narrow H$\beta$ luminosity as:
\begin{equation}\label{ext_corr}
L_{\rm nH\beta,corr} = \left(\frac{L_{\rm nH\alpha}/L_{\rm nH\beta}}{3.1}\right)^{3.37} \times L_{\rm nH\beta}
\end{equation}
\citep[see Appendix D of][for a derivation]{lamassa2023}. This correction is only applied if the observed ratio between nH$\alpha$ and nH$\beta$ exceeds 3.1.

If we were to use the $\lambda^{-0.7}$ extinction law instead, this would change both the normalization in Equation \ref{lbol} (from -0.25 to -0.52) and the form of the reddening correction to $L_{\rm nH\beta,corr} = \left(\frac{L_{\rm nH\alpha}/L_{\rm nH\beta}}{3.1}\right)^{5.28} \times L_{\rm nH\beta}$. For the BASS AGN, the \citet{cardelli1989} extinction law produces a $L_{\rm bol}$ value that ranges from 0.14 - 1.86 times the $L_{\rm bol}$ value found using the $\lambda^{-0.7}$ extinction law, with a median diference of 1.28. Propagating these differences into Equations \ref{l5100_opt} and \ref{o3_hb_bhmass_formula}, the choice of the optical attenuation law can affect the derived black hole mass by a factor of 0.31 - 1.45, with a median difference of 1.16.

\subsubsection{$\lambda L_{\rm 5100\angstrom,X-ray}$ : Estimating $\lambda L_{\rm 5100,\angstrom}$ from Intrinsic X-ray Luminosity}

There is a connection between the ultraviolet emission (2500 $\angstrom$) from the accretion disk and soft X-ray emission (2 keV) from the corona. This relationship is parameterized by $\alpha_{\rm OX}$ which is the slope of a powerlaw spectrum that connects the optical to X-ray emission \citep{tananbaum1979}. This relationship can be used to estimate the optical continuum luminosity from the intrinsic X-ray luminosity \citep[see, e.g.,][]{maiolino2007}. Using a sample of X-ray selected Type 1 AGN from the {\it XMM}-COSMOS survey \citep{hasinger2007,cappelluti2009}, \citet{lusso2010} quantified a relationship between the monochromatic X-ray luminosity at 2 keV and ultraviolet continuum luminosity at 2500 $\angstrom$. Starting from this relationship, we derive an equation that links the intrinsic integrated 2-10 keV luminosity to the 5100 $\angstrom$ continuum luminosity (see the Appendix for the derivation):

\begin{equation}
  \begin{split}\label{l5100_xray}
    {\rm \log}\left(\frac{\lambda L_{\rm 5100 \angstrom,X-ray}}{10^{44} \, {\rm erg \, s^{-1}}}\right) = 1.316 \, \times \\
    \rm{\log}\left(\frac{L_{\rm 2-10keV,intrinsic}}{10^{43} \, {\rm erg \, s^{-1}}}\right) - 1.378
  \end{split}
\end{equation}

\subsubsection{Comparing Proxies of $\lambda L_{\rm 5100,\angstrom}$}
\begin{figure}
 \includegraphics[scale=0.52]{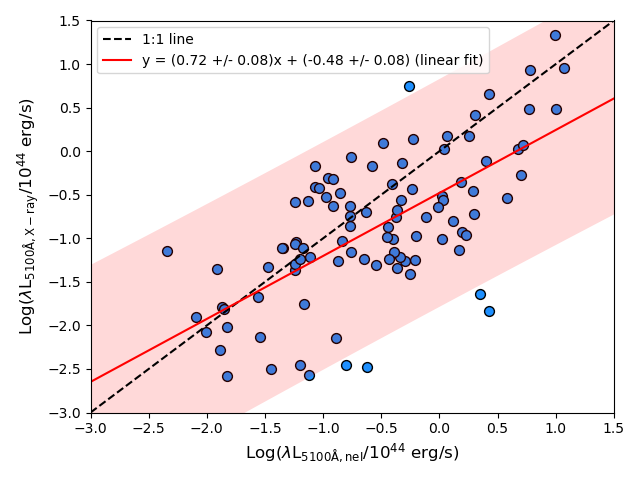}
 \caption{\label{l5100_comp} Comparison of estimates of the accretion disk luminosity at 5100 $\angstrom$ ($\lambda L_{\rm 5100\angstrom}$) using optical narrow emission lines ($\lambda L_{\rm 5100\angstrom,nel}$) and the intrinsic 2-10 keV luminosity ($\lambda L_{\rm 5100\angstrom,X-ray}$) as proxies. The black dashed line indicates the one-to-one relation, the red solid line is a linear fit to the data, and the shaded red region shows the 95\% prediction interval for the linear fit. The mean offset (i.e., $\log$($L_{\rm 5100\angstrom,X-ray}$) - $\log$($L_{\rm 5100\angstrom,nel}$)) is -0.32 $\pm$ 0.68 dex. While these quantities are statistically equal, this scatter of $\sim$5 in $\lambda L_{\rm 5100\angstrom}$ can propagate to uncertainties of $\sim$2 in black hole mass estimates when using Equation \ref{o3_hb_bhmass_formula}. Error bars are not shown here since they are generally smaller than the size of the symbols.}
\end{figure}

In Figure \ref{l5100_comp}, we compare $\lambda L_{\rm 5100 \angstrom,nel}$ with $\lambda L_{\rm 5100 \angstrom,X-ray}$ for the 99 AGN for which there is only a narrow H$\alpha$ component detected and we can thus use the Balmer decrement to calculate $L_{\rm nH\beta,corr}$ and then estimate $\lambda L_{\rm 5100 \angstrom,nel}$. As summarized in Table \ref{comparison_summary}, we find a mean offset  (i.e., $\log$($\lambda L_{\rm 5100\angstrom,X-ray}$) - $\log$($\lambda L_{\rm 5100\angstrom,optical}$) of -0.32 dex with a large scatter of 0.68 dex. Though there is an average offset, the wide scatter indicates that the quantities are statistically equal, though the scatter of up to a factor of 5 propagates into black hole mass uncertainties up to a factor of 2 (see Equation \ref{o3_hb_bhmass_formula}). For completeness, we quantify the relationship between $\lambda L_{\rm 5100 \angstrom,nel}$ and $\lambda L_{\rm 5100 \angstrom,X-ray}$ by using a linear least square fitter (\textsc{LinearLSQFitter} in astropy). Though we find a linear relationship above the one-to-one line, the 95\% prediction interval to the linear fit overlaps the one-to-one relation due to the wide scatter in the fit residuals ($\sigma_{\rm residual}$).

\begin{deluxetable*}{llllllllll}
\tablecaption{\label{comparison_summary}Comparison of $\lambda L_{\rm5100\angstrom}$ Proxies and Black Hole Masses}
\tablehead{
 \colhead{$x$} & \colhead{$y$} &  \colhead{Mean Offset\tablenotemark{a}} & \colhead{Dispersion\tablenotemark{a}} & \colhead{$R$\tablenotemark{b}} & \colhead{$p$\tablenotemark{b}} & \colhead{Slope\tablenotemark{c}} & \colhead{Intercept\tablenotemark{c}} & \colhead{$\sigma_{\rm residual}$\tablenotemark{d}} &\colhead{$n$\tablenotemark{e}}\\
  & & \colhead{(dex)} & \colhead{(dex)} }
\startdata
$\lambda L_{\rm 5100 \angstrom,nel}$  & $\lambda L_{\rm 5100 \angstrom,X-ray}$  & -0.32 & 0.68 & 0.6565 & 1.6$\times10^{-13}$ & 0.72$\pm$0.08 & -0.48$\pm$0.08 & 0.65 & 99 \\
$M_{\rm BH,Pa\alpha}$ & $M_{\rm [O\,III]/nH\beta}$ & 0.39 & 0.44 & 0.7105 & 0.004 & 0.49$\pm$0.14 & 4.42$\pm$1.11 & 0.32 & 14 \\
$M_{\rm BH,Pa\beta}$ & $M_{\rm [O\,III]/nH\beta}$ & 0.48 & 0.51 & 0.6850 & 0.014 & 0.57$\pm$0.19 & 3.71$\pm$1.47 & 0.45 & 12 \\
$M_{\rm BH,\sigma_*}$ & $M_{\rm BH,Pa\alpha}$ & 0.08 & 0.33 & 0.8694 & 0.0002 & 0.93$\pm$0.17 & 0.63$\pm$1.30 & 0.36 & 12 \\
$M_{\rm BH,\sigma_*}$ & $M_{\rm BH,Pa\beta}$ & -0.38 & 0.35 & 0.8719 & 0.0010 & 0.95$\pm$0.19 & 0.02$\pm$1.49 & 0.39 & 10 \\
$M_{\rm BH,\sigma_*}$ & $M_{\rm [O\,III]/nH\beta}$ & -0.08 & 0.74 & 0.3327 & 3.0$\times10^{-5}$ & 0.40$\pm$0.09 & 4.78$\pm$0.75 & 0.66 & 151 \\
\enddata
\tablenotetext{a}{The mean offset and dispersion is calculated from $\log$($y$) - $\log$ ($x$). These results are equivalent to fitting a line between these parameters where the slope is fixed to unity and the offset represents the $y$-intercept.}
\tablenotetext{b}{$R$ is the Pearson correlation coefficient and $p$ is the $p$-value to measure the significance of a correlation. A $p$-value $<$ 0.05 indicates a significant ($\geq$ 2$\sigma$) correlation. Quoted results are from the \textsc{pearsonr} function in \textsc{scipy}. }
\tablenotetext{c}{Fitted slope and intercept from a linear fit to the quantities using \textsc{LinearLSQFitter} in astropy.}
\tablenotetext{d}{Standard deviation in the fit residuals, calculated using $ \sqrt{ \sum_{i=1}^{n} \frac{(y_i - \hat y)^2}{(n-2)}}$, where $y_i$ is the measured value, $\hat y$ is the predicted value from the fitted relation, and $n$ is the number of data points.}
\tablenotetext{e}{Number of sources in sample.}
\end{deluxetable*}

Since we have intrinsic X-ray luminosity measurements for a larger number of AGN than those that have only narrow H$\alpha$ emission detected, we use $\lambda L_{\rm 5100\angstrom,X-ray}$ as $\lambda L_{\rm 5100\angstrom}$, together with Equation \ref{hafwhm_eq}, as input into Equation \ref{o3_hb_bhmass_formula}.  From this formula, we calculate the $L_{\rm [O\,III]}/L_{\rm nH\beta}$-derived black hole masses ($M_{\rm BH,[O\,III]/nH\beta}$) and report these values in Table \ref{bass_mbh_table}.

\begin{deluxetable*}{lllll}
\tablecaption{\label{bass_mbh_table}BASS AGN Black Hole Masses\tablenotemark{1}}
\tablehead{
  \colhead{ID} & \colhead{$\log$($M_{\rm [O\,III]/nH\beta}$)} & \colhead{$\log$($M_{\rm BH,Pa\alpha}$)} & \colhead{$\log$($M_{\rm BH,Pa\beta}$)} & \colhead{$\log$($M_{\rm BH,\sigma_*}$)} \\
 & \colhead{M$_{\rm \odot}$} & \colhead{M$_{\rm \odot}$} & \colhead{M$_{\rm \odot}$} & \colhead{M$_{\rm \odot}$}}
\startdata
63 & 6.80 & \nodata & 6.46 & 7.29 \\
72 & 8.73 & 8.19 & 8.03 & 8.06 \\
226 & 8.13 & 7.94 & 7.88 & \nodata \\
404 & 8.17 & 7.21 & \nodata & 6.86 \\
511 & 8.16 & 7.75 & 6.99 & 7.56 \\
586 & 8.14 & 7.66 & \nodata & 8.14 \\
682 & 7.56 & \nodata & 7.89 & 8.71 \\
698 & 8.65 & 8.35 & \nodata & 8.3 \\
700 & 8.90 & 8.13 & \nodata & 7.9 \\
738 & 7.68 & 7.34 & \nodata & 7.37 \\
757 & 9.03 & 8.54 & 8.67 & 8.4 \\
971 & 8.65 & 8.94 & \nodata & 8.12 \\
1027 & 7.96 & \nodata & 7.46 & 7.85 \\
1060 & 7.94 & 8.3 & 8.02 & 8.35 \\
1090 & 8.00 & 8.22 & 8.04 & \nodata \\
1138 & 8.68 & \nodata & 7.57 & 8.34 \\
1157 & 7.58 & 6.41 & 6.2 & 6.3 \\
1161 & 8.14 & 7.47 & 7.73 & 7.93 \\
\enddata
\tablenotetext{1}{$M_{\rm [O\,III]/nH\beta}$, $M_{\rm BH,Pa\alpha}$, and and $M_{\rm BH,Pa\beta}$ are calculated here using emission line fluxes from \citet{oh2022,lamperti2017,riccif2022,denbrok2022} and the intrinsic 2-10 keV X-ray flux from \citet{riccic2017}. Black hole masses calculated from the stellar velocity dispersions ($M_{\rm BH,\sigma*}$) reported in \citet{koss2022b}. We show here a subset of this table for illustrative purposes. The catalog is available in its entirety on-line.}
\end{deluxetable*}

\subsection{Correlation between NLR Ionization Hardness and NIR BLR Kinematics?}
The utility of $\log$($L_{\rm [O\,III]}/L_{\rm nH\beta}$) as a parameter to estimate black hole mass relies on its observed correlation with the FWHM of H$\alpha$, which traces the virial motion of gas in the broad line region. \citet{kim2010} demonstrated that the FWHMs of the NIR Paschen lines Pa$\alpha$ and Pa$\beta$ are well correlated with the FWHM of the optical Balmer lines, though the Paschen lines have systematically lower FWHM values compared with the Balmer lines \citep[though see][who find no such offset]{riccif2017}. \citet{kim2010} attribute this correlation to the Paschen lines also forming within the BLR but at a further distance from the black hole compared with the Balmer lines due to ionization stratification where the gas producing the Paschen lines is orbiting at a lower speed.  Assuming the motion of the gas producing the Paschen lines is dominated by its orbital velocity around the SMBH, virial black hole mass formulas based on the FWHM of Paschen lines and either luminosity of these emission lines or NIR continuum luminosity can be used to estimate $M_{\rm BH}$ \citep{kim2010,landt2011a,landt2011b,landt2013}.

\begin{figure*}
\includegraphics[scale=0.55]{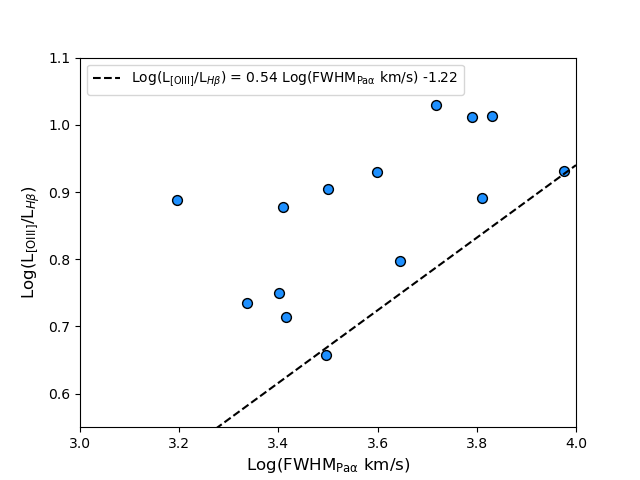}~
\includegraphics[scale=0.55]{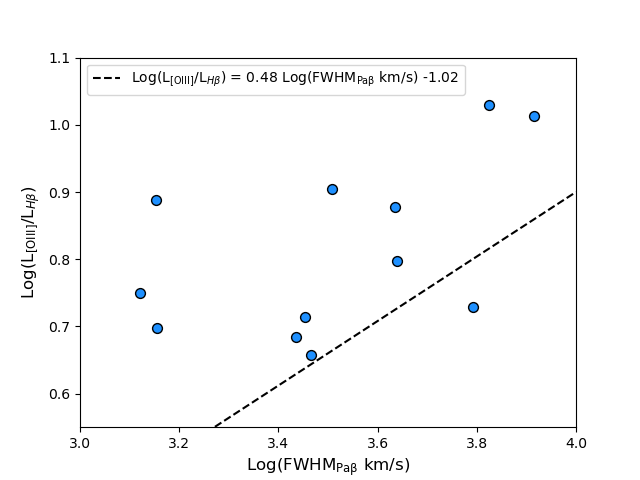}
\caption{\label{o3hb_fwhmpa} Comparison of narrow line region ionization field hardness ($\log$($L_{\rm [O\,III]}/L_{\rm nH\beta}$)) and broad line region kinematics traced by the FWHM of the broad Pa$\alpha$ (left) and Pa$\beta$ (right) lines. We find a significant correlation between NLR ionization field hardness and FWHM$_{\rm Pa\alpha}$ ($p$-value = 0.0297) but an insignificant correlation with FWHM$_{\rm Pa\beta}$ ($p$-value=0.0918). For reference, the expected relationship between $\log$($L_{\rm [O\,III]}/L_{\rm nH\beta}$) and $\log$(FWHM$_{\rm Pa\alpha}$) and $\log$(FWHM$_{\rm Pa\beta}$), combining the results of \citet{kim2010} and BM19 (Eqs. \ref{o3hb_pa_fwhmha} and \ref{o3hb_pb_fwhmha}), is shown by the dashed black line. The correlation between ionizaton field hardness and BLR kinematics is weaker than that reported in BM19. Either such trends are only prominent when considering aggregate AGN spectra (as in BM19) and not individual measurements, the coupling between the NLR ionization field and the more distant BLR in which the Paschen lines form is weaker, or extinction in the BLR that extinguishes broad H$\alpha$ in these AGN also depresses the Paschen emission relative to the unobscured AGN studied in BM19. Error bars are not shown here since they are generally smaller than the size of the symbols.}
  \end{figure*}

Here we test whether there is a correlation between $\log$($L_{\rm [O\,III]}/L_{\rm nH\beta}$) and the FWHM of Pa$\alpha$ and Pa$\beta$ such that we may expect whether or not there would be agreement in the black hole masses derived using Equation \ref{o3_hb_bhmass_formula} and the Paschen virial black hole mass equations. We plot these quantities in Figure \ref{o3hb_fwhmpa} and find a significant correlation between $\log$($L_{\rm [O\,III]}/L_{\rm nH\beta}$) and $\log$(FWHM$_{\rm Pa\alpha}$) (Pearson correlation coefficient = 0.58, with $p$-value = 0.0297) though an insignificant correlation between $\log$($L_{\rm [O\,III]}/L_{\rm nH\beta}$) and $\log$(FWHM$_{\rm Pa\beta}$) (Pearson correlation coefficient = 0.508, $p$-value = 0.0918).

For comparison, we also include a line marking the expected relationship between $\log$($L_{\rm [O\,III]}/L_{\rm nH\beta}$) and $\log$(FWHM$_{\rm Pa\alpha}$) and $\log$(FWHM$_{\rm Pa\beta}$) using Equation 1 from BM19 and the relationships found between the H$\alpha$ FWHM and Pa$\alpha$ and Pa$\beta$ FWHM from \citet{kim2010}:

\begin{equation}
  \begin{split}
    {\rm\log}\left(\frac{{\rm FWHM_{\rm H\alpha}}}{{\rm 1000 \, km \, s^{-1}}}\right) = (0.934 \pm 0.084) \, \times \\
    \rm{\log}\left(\frac{{\rm FWHM_{\rm Pa\alpha}}}{{\rm 1000 \, km \, s^{-1}}}\right) + (0.074 \pm 0.038),
  \end{split}
\end{equation}

\begin{equation}
  \begin{split}
    {\rm \log}\left(\frac{{\rm FWHM_{\rm H\alpha}}}{{\rm 1000 \, km \, s^{-1}}}\right) = (0.821 \pm 0.075) \, \times \\
    \rm{\log}\left(\frac{{\rm FWHM_{\rm Pa\beta}}}{{\rm 1000 \, km \, s^{-1}}}\right) + (0.076 \pm 0.038),
  \end{split}
\end{equation}

Combining these equations with Equation 1 from BM19, we have:

\begin{equation}\label{o3hb_pa_fwhmha}
  \begin{split}
    {\rm \log}\left(\frac{L_{\rm [OIII]}}{L_{\rm nH\beta}}\right) = (0.54 \pm 0.08) \, \times \\
    \rm{\log}\left(\frac{{\rm FWHM_{\rm Pa\alpha}}}{{\rm km \, s^{-1}}}\right) - (1.22 \pm 0.73),
  \end{split}
\end{equation}

\begin{equation}\label{o3hb_pb_fwhmha}
  \begin{split}
    {\rm \log}\left(\frac{L_{\rm [OIII]}}{L_{\rm nH\beta}}\right) = (0.48 \pm 0.07) \, \times \\
    \rm{\log}\left(\frac{{\rm FWHM_{\rm Pa\beta}}}{{\rm km \, s^{-1}}}\right) - (1.02 \pm 0.60),
  \end{split}
\end{equation}

Though there is wide scatter in both relationships, it is striking that the measured $\log$($L_{\rm [O\,III]}/L_{\rm nH\beta}$) values are systematically higher than the predicted trend line shown in Figure \ref{o3hb_fwhmpa}. Comparing these results with Figure 4 of BM19, which plots the relationship between $\log$($L_{\rm [O\,III]}/L_{\rm nH\beta}$) and $\log$(FWHM$_{\rm bH\alpha}$), we see that the measurements of $\log$($L_{\rm [O\,III]}/L_{\rm nH\beta}$) from individual spectra are systematically higher than those derived from the stacked spectra and their derived trend line, similar to what we see in Figure \ref{o3hb_fwhmpa}. BM19 acknowledge that there is a weaker relationship between $\log$($L_{\rm [O\,III]}/L_{\rm nH\beta}$) and  $\log$(FWHM$_{\rm bH\alpha}$) when considering line measurements from individual spectra compared with the median spectra from which they derived their Equation 1. They attribute this weaker correlation, and systematically higher ($L_{\rm [O\,III]}/L_{\rm nH\beta}$) ratio, to uncertainties in decomposing the narrow H$\beta$ line from the broad line, which underestimates the narrow H$\beta$ flux.

As our sample is not afflicted by similar uncertaintites in the H$\beta$ line decomposition, we conclude that the correlation between ionization field hardness and the FWHM of the broad Paschen lines is weaker than that reported for the broad H$\alpha$ line. It could be that trends between NLR ionization hardness and BLR kinematics are stronger when considering the AGN population in aggregrate, as was done in BM19, rather than individually. Or there may be a physical driver: if the BLR is ionization stratified and the Paschen lines form further out from the center of the accretion disk than the Balmer lines, then perhaps there is a weaker coupling between the NLR ionization hardness and outer BLR compared with the inner BLR. Additionally, the AGN in this analysis are moderately obscured while BM19 analyzed Type 1, unobscured AGN: extinction towards the BLR may depress the broad Paschen lines causing the narrow [\ion{O}{3}]/H$\beta$ line ratio to appear relatively enhanced.

\subsection{Comparing $L_{\rm [O\,III]}/L_{\rm nH\beta}$-derived $M_{\rm BH}$ with Paschen-derived $M_{\rm BH}$}

We use the following formulas from \citet{denbrok2022}, originally published in \citet{kim2010} but assuming a virial factor of $f=1$, to calculate black hole masses from the NIR Paschen lines:
\begin{equation}
  \begin{split}
    {\rm log} \left(\frac{M_{\rm BH, Pa\alpha}}{M_{\odot}}\right) = 7.16 + 0.43 \times {\rm \log} \left(\frac{L_{\rm Pa \alpha}}{10^{42} \, {\rm erg \, s^{-1}}} \right) \\
    + 1.92 \times {\rm \log} \left(\frac{{\rm FWHM_{\rm Pa\alpha}}}{10^{3} \, {\rm km \, s^{-1}}}\right)
    \end{split}
\end{equation}
\begin{equation}
  \begin{split}
    {\rm \log} \left(\frac{M_{\rm BH, Pa\beta}}{M_{\odot}}\right) = 7.20 + 0.45 \times {\rm \log} \left(\frac{L_{\rm Pa \beta}}{10^{42} \, {\rm erg \, s^{-1}}} \right) \\
   + 1.69 \times {\rm \log} \left(\frac{{\rm FWHM_{\rm Pa\beta}}}{10^{3} \, {\rm km \, s^{-1}}}\right).
    \end{split}
\end{equation}
In Table \ref{bass_mbh_table}, we list the black hole masses calculated from these formulas for the AGN that have reliable fits to the broad Pa$\alpha$ or Pa$\beta$ lines.

\begin{figure*}
\includegraphics[scale=0.55]{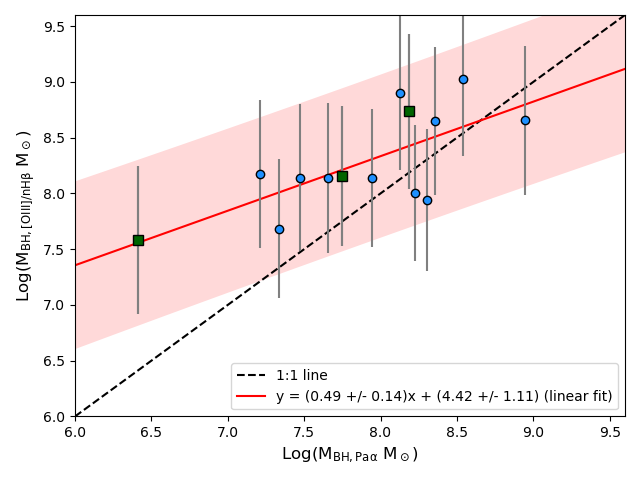}~
\includegraphics[scale=0.55]{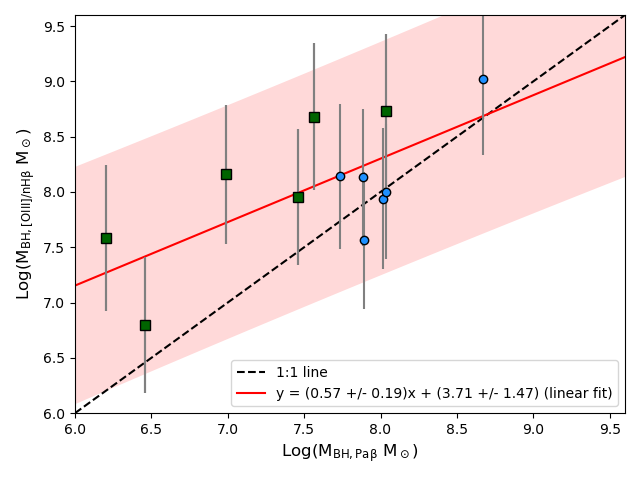}
\caption{\label{mbho3hb_mbhpa} Comparison of black hole masses calculated using the $L_{\rm [O\,III]}/L_{\rm nH\beta}$ method in Equation \ref{o3_hb_bhmass_formula} ($M_{\rm BH,[O\,III]/nH\beta}$) and those derived from the broad Pa$\alpha$ line (left) and broad Pa$\beta$ line (right).The dark green squares indicate AGN whose Paschen emission line parameters were derived from the \citet{denbrok2022} catalog where the quality of the spectral fit is not reported. $M_{\rm BH,[O\,III]/nH\beta}$ is systematically larger than masses derived via the broad Paschen lines by a factor of $\sim2-3$, though the scatter is slightly larger than these offsets (see Table \ref{comparison_summary}). The agreements between the Paschen-derived black hole masses and those from the $L_{\rm [O\,III]}/L_{\rm nH\beta}$ method is better when censoring the \citet{denbrok2022} data (with mean offsets of 0.30$\pm$ 0.42 dex and 0.09$\pm$0.26 dex when comparing  $M_{\rm BH,[O\,III]/nHn\beta}$ with $M_{\rm BH,Pa\alpha}$ and $M_{\rm BH,Pa\beta}$, respectively), but small number statistics preclude us from drawing conclusions from this censored dataset. Error bars are derived by propagating the 0.68 dex uncertainty on $\lambda L_{\rm 5100\angstrom}$ and the uncertainties on FWHM$_{\rm bH\alpha}$ (from the errors on the slope and normalization in Equation \ref{hafwhm_eq}) to the $M_{\rm BH,[O\,III]/nH\beta}$ equation (Equation \ref{o3_hb_bhmass_formula}).}
\end{figure*}

In Figure \ref{mbho3hb_mbhpa}, we compare $M_{\rm BH,[O\,III]/nHn\beta}$ with those calculated from the Paschen lines using the equations above, acknowledging that the marginal correlation found between $\log$($L_{\rm [O\,III]}/L_{\rm nH\beta}$) and FWHM$_{\rm Pa\beta}$ suggests that the relationship between the black hole masses calculated from these quantities may show poor agreement. Consistent with those expectations, we find a worse agreement between $M_{\rm BH,[O\,III]/nH\beta}$ and $M_{\rm BH,Pa\beta}$ ($p$ = 0.014) than between  $M_{\rm BH,[O\,III]/nH\beta}$ and $M_{\rm BH,Pa\alpha}$ ($p$=0.004). $M_{\rm BH,[O\,III]/nH\beta}$ is systematically higher than the black hole masses derived from the Paschen lines, by an average factor of 0.39 $\pm$ 0.44 dex and 0.48 $\pm$ 0.51 dex for $M_{\rm BH,Pa\alpha}$ and $M_{\rm BH,Pa\beta}$, respectively. In both cases, we find a fitted slope that is shallower than unity.  However, the residuals we find when fitting a linear relationship to the black hole mass proxies is consistent with the residuals when assuming a fixed slope of unity (see Table \ref{comparison_summary}).

Since \citet{denbrok2022} do not report the quality of their spectral fits while the other two BASS near-infrared catalogs we queried did \citep{lamperti2017,riccif2022}, we investigate the impact that censoring these data have on our results. In Figure \ref{mbho3hb_mbhpa}, we mark the AGN with \citet{denbrok2022} Paschen emission line measurements with dark green squares. When we remove these datapoints and repeat the analysis above, we find a better average agreement between $M_{\rm BH,[O\,III]/nH\beta}$ and $M_{\rm BH,Pa\alpha}$ (average offset of 0.30$\pm$0.42 dex) and much closer agreement between $M_{\rm BH,[O\,III]/nH\beta}$ and $M_{\rm BH,Pa\beta}$ (average offset of 0.09$\pm$0.26 dex), but this censoring only leaves us with 6 AGN in the Pa$\beta$ sample.

Though the BASS AGN sample represents one of the most comprehensive AGN datasets, boasting rich optical, NIR, and X-ray spectroscopy \citep{oh2022,koss2022b,riccif2017,lamperti2017,denbrok2022,riccic2017}, this analysis is still statistically limited with only 14 (12) AGN that have measured broad Pa$\alpha$ (Pa$\beta$) emission. However, these results allow us to comment on trends and provide context for studies that use the BM19 method to estimate black hole mass \citep{rey2021,ferre-mateu2021,vietri2022,siudek2023} and as an observational benchmark for theoretical simulations \citep{shankar2020,volenteri2020,habouzit2021,dubois2021,treibitsch2021,trinca2022,sassano2023,beckmann2023}.

It is typical to find a factor of a few spread in black hole mass measurements using different methods. For instance, mass measurements using virial black hole mass formulas from the Balmer lines can be discrepant with direct mass measurements (i.e., measuring gas or stellar dynamics, reverberation mapping) by a factor of 0.3-0.5 dex \citep[e.g.,][]{peterson2014}. At higher redshift, the uncertanities in black hole masses derived via reverberation mapping campaigns and single epoch virial mass formuals are as high as 0.45 dex for \ion{Mg}{2} and 0.58 for \ion{C}{4}. A similar comparison to what we have presented here is to compare black holes masses calculated from single epoch virial mass formulas. Here too, comparisons between $M_{\rm BH}$ derived from \ion{C}{4}, \ion{Mg}{2}, H$\beta$, and H$\alpha$ show a spread of $\sim$0.12-0.40 dex \citep{trakhtenbrot2012,shen2012}.

By comparison, the dispersion in $\log$($M_{\rm BH,[O\,III]/nH\beta}$) - $\log$($M_{\rm BH,Pa\alpha}$) is on the high tail of the distribution found from intercomparisons of optical and ultraviolet broad line virial mass measurements though similar to the dispersion found when comparing black holes masses calculated from reverberation mapping with those measured from single epoch spectroscopy. However, the systematic offset between $M_{\rm BH,[O\,III]/nH\beta}$ and $M_{\rm BH,Pa\alpha}$ is higher than what is typically observed in these intercomparisons ($\sim$0.05 - 0.2 dex). The agreement of Paschen-derived black hole masses with those calculated from reverberation mapping or Balmer line virial mass formulas is also tighter \citep[$\sim$0.2 dex][]{kim2010} than we see in our comparison between $M_{\rm BH,[O\,III]/nH\beta}$ and the Paschen-derived black hole masses.

\subsection{Masses derived from the stellar velocity dispersion}
The well known correlation between the central supermassive black hole mass and the velocity dispersion of stars ($\sigma_{\rm *}$) in the host galaxy \citep[e.g.,][]{gebhardt2000,ferrarese2000} offers a method for calculating black hole masses from $\sigma_{\rm *}$ \citep[$M_{\rm BH,\sigma_*}$;][]{ferrarese2001}. \citet{koss2022b} measured the stellar velocity dispersions of 484 BASS AGN by using the penalized \textsc{pPXF} code \citep{cappellari2004} to fit a model galaxy spectrum (using high resolution galaxy templates) to the spectra of the BASS AGN host galaxies, observed mostly with VLT/X-Shooter and Palomar/Double Spec. \citet{koss2022b} used \textsc{pPXF} to calculate the velocity dispersions from the \ion{Ca}{2} H and K $\lambda$3969, 3934, \ion{Mg}{1} $\lambda$5175, and \ion{Ca}{2} triplet $\lambda$8498, 8542, 8662 absorption lines \citep[see][for full details]{koss2022b}. From the velocity dispersions, they calculated black hole masses using this relation from \citet{kormendy2013}:
\begin{equation}
\rm{log}\left(\frac{M_{\rm BH,\sigma_*}}{M_{\rm \odot}}\right) = 4.38 \, \times \, \rm{log}\left(\frac{\sigma_{*}}{200 \, km\,s^{-1}}\right) + 8.49
\end{equation}
We report $M_{\rm BH,\sigma_*}$ for the BASS AGN used in this analysis in Table \ref{bass_mbh_table}.

\begin{figure*}
\includegraphics[scale=0.55]{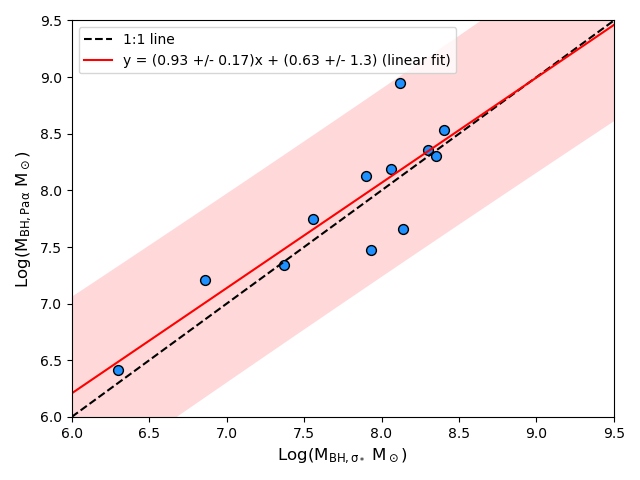}~
\includegraphics[scale=0.55]{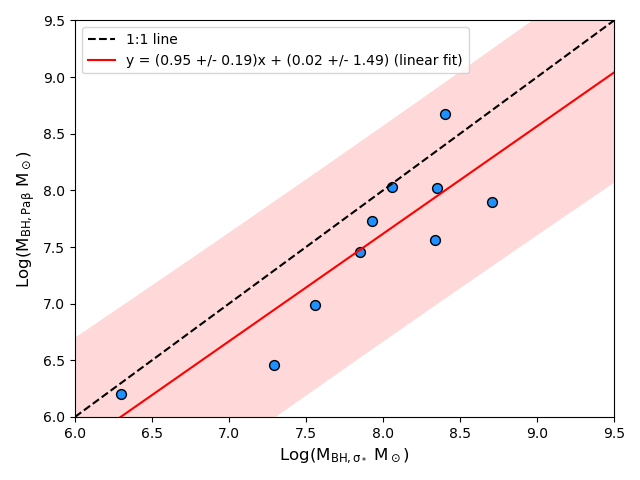}
\caption{\label{mbhvdisp_mbhpa} Comparison of black hole masses calculated from the host stellar velocity dispersion ($M_{\rm BH,\sigma_*}$) and from the broad Pa$\alpha$ line (left) and broad Pa$\beta$ line (right). While there is a good agreement with $M_{\rm BH,Pa\alpha}$ (average offset of -0.08 dex with standard deviation of 0.33 dex), $M_{\rm BH,\sigma_*}$ is systematically higher than $M_{\rm BH,Pa\beta}$ by a factor of 0.38 $\pm$ 0.35 dex, though the quanties are consistent with the one-to-one relation at the 95\% prediction level. Line styles are the same as Figure \ref{l5100_comp}.}
\end{figure*}

First, we compare the stellar velocity dispersion-derived black hole masses with those calculated from the Paschen lines to test whether any systematic offsets are seen that are similar to what we observed in our $M_{\rm BH,[O\,III]/nH\beta}$ analysis above. As we show in Figure \ref{mbhvdisp_mbhpa} and report in Table \ref{comparison_summary}, a systematic offset is seen between $M_{\rm BH,\sigma_*}$ and the Pa$\beta$-derived black hole mass (-0.38 dex with a dispersion of 0.35 dex), though there is much better agreement with the Pa$\alpha$-derived black hole mass (0.08 dex with a dispersion of 0.33 dex). These results, when considered with the analysis above, indicate that the Pa$\beta$-derived black hole masses used in this analysis may be biased low.

\begin{figure}
 \includegraphics[scale=0.55]{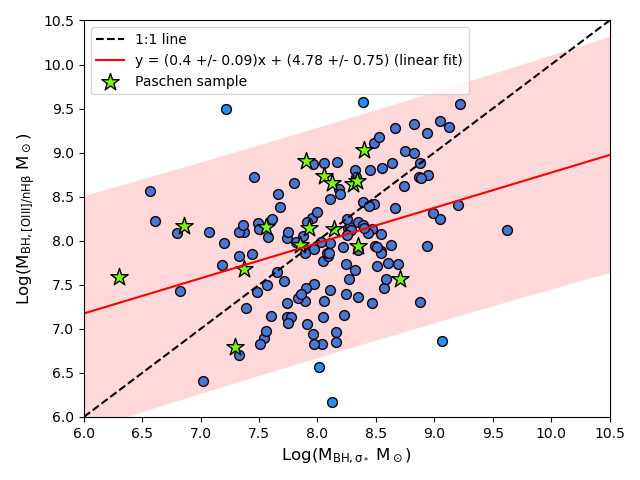}~
 \caption{\label{mbho3hb_mbhvdisp} {\it Left}: Comparison of $L_{\rm [O\,III]}/L_{\rm nH\beta}$-derived black hole masses ($M_{\rm BH,[O\,III]/nH\beta}$) and those calculated from the host stellar velocity dispersion ($M_{\rm BH,\sigma_*}$). For reference, the AGN we used to assess the agreement between $M_{\rm BH,[O\,III]/nH\beta}$ and the Paschen-derived black hole masses are shown by the green stars. On average, the values are consistent to within 0.08 dex, albeit with a wide scatter (0.74 dex). Line styles are the same as Figure \ref{l5100_comp}.}
\end{figure}

We compare the $L_{\rm [O\,III]}/L_{\rm nH\beta}$-derived black hole masses with the stellar velocity dispersion-derived black hole masses in Figure \ref{mbho3hb_mbhvdisp}. There is a wide dispersion between values (0.74 dex), but these two methods of calculating black hole masses are in much better agreement (average $\log$($M_{\rm BH,[O\,III]/nH\beta}$) - $\log$($M_{\rm BH,\sigma_*}$) = -0.08 dex) than when comparing $M_{\rm BH,[O\,III]/nH\beta}$  with the Paschen-derived black hole masses.

BM19 also compared $M_{\rm BH,[O\,III]/nH\beta}$ with the stellar velocity dispersion and stellar mass of the AGN host galaxies for the 10,000 Type 2 SDSS AGN they analyzed. They performed a maximum likelihehood estimation to quantify the relationship between $M_{\rm BH,[O\,III]/nH\beta}$ and $\sigma_{\rm *}$, finding a scatter of 0.45 dex (standard deviation). They thus conclude that their method provides black hole mass estimates as accurate as those derived from the $M_{\rm BH} - \sigma{\rm *}$ relation, though they stress that a direct test would come by comparing $M_{\rm BH,[O\,III]/nH\beta}$ with those derived from the broad Paschen lines. 

Finally we note that two of the BASS AGN we analyzed here have black hole masses measured via water megamaser disks \citep{greene2016}: NGC 1194 and Circinus. We find a consistent $M_{\rm BH}$ value between the megamaser method and the BM19 method for NGC 1194 (7.85 dex compared with 7.95 dex, respectively), but a disagreement of almost 2 dex between these methods for Circinus (6.06 dex compared with 7.95 dex, respectively). The disagreement for Circinus could be due to its much closer proximity ($z = 0.0015$) compared with the typical SDSS AGN from which the BM19 was derived ($0.01 < z < 0.3$): the optical spectra from Circinus was observed with the Very Large Telescope using a 2$^{\prime\prime}$ wide spectral slit \citep{koss2022c}, so only the inner 60 pc of this galaxy was sampled, while the size scales probed by the SDSS AGN are 0.6 - 1.2 kpc. If the $L_{\rm [O\,III]}/L_{\rm nH\beta}$ ratio changes appreciably from circumnuclear scales (tens of parsecs) to extended scales (hundreds of parsecs), then the relationship calibrated using data extracted from a larger region may not be appropriate when applied to much smaller physical scales. For instance, SDSS MaNGA observations of nearby AGN demonstrate that $L_{\rm [O\,III]}/L_{\rm nH\beta}$ can sometimes be elevated closer to the black hole compared with extended scales \citep[e.g.][]{alban2023} which would boost the $M_{\rm BH}$ value calculated from Equation \ref{o3_hb_bhmass_formula} when measuring $L_{\rm [O\,III]}/L_{\rm nH\beta}$ from a compact region.

\subsection{Properties of Obscured AGN in BASS}

\citet{koss2022b} reported the distribution of black hole masses (calculated from $\sigma_{\rm *}$) and Eddington ratios ($L_{\rm bol}/L_{\rm Edd}$, where $L_{\rm bol}$ is the AGN bolometric luminosity and $L_{\rm Edd}$ is the Eddington luminosity, $L_{\rm Edd} = 1.26 \times 10^{38} M_{\rm BH}/M_{\rm \odot}$) of obscured AGN (Type 1.9 and Type 2) from the BASS survey. Using the $M_{\rm BH,[O\,III]/nH\beta}$ values we calculated, we compare the distribution of $M_{\rm BH}$ and Eddington ratios of this sample with the larger BASS sample from \citet{koss2022b}. To be consistent with the methodology of \citet{koss2022b}, we calculate $L_{\rm bol}$ from the intrinsic 14-150 keV luminosity reported in \citet{riccic2017} and use a bolometric correction factor of 8 \citep{vasudevan2009}. Using the $\sigma_{\rm *}$ values reported in \citet{koss2022b} and X-ray luminosities reported in \citet{riccic2017}, we calculate $M_{\rm BH,\sigma_*}$ and the Eddington ratio for 323 obscured AGN from the BASS sample.

\begin{figure*}
\includegraphics[scale=0.55]{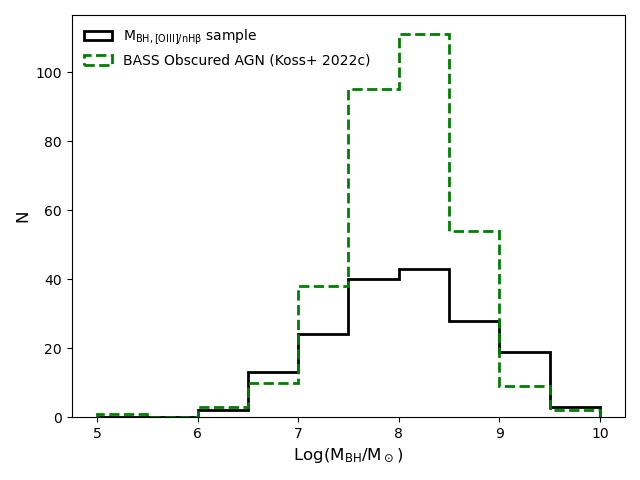}~
\includegraphics[scale=0.55]{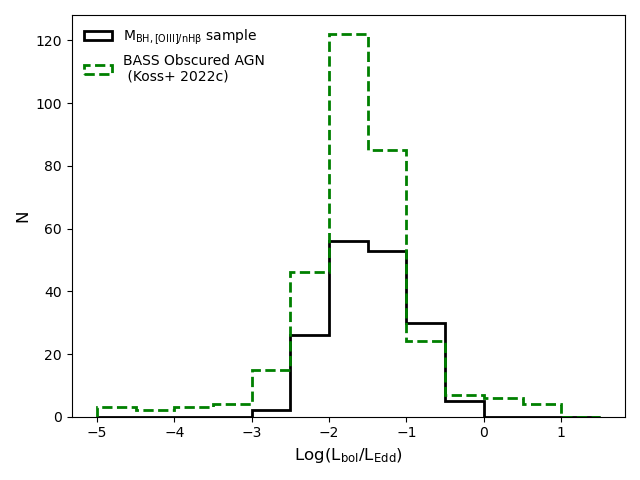}
\caption{\label{bass_comp} Comparison of $M_{\rm BH,[O\,III]/nH\beta}$ values from this study (black solid line) and $M_{\rm BH}$ derived from $\sigma_{\rm *}$  for a sample of 323 obscured BASS AGN \citep[green dashed line;][]{koss2022c}. The distributions are similar with median black hole mass values of $\log$($M_{\rm BH}/L_{\rm \odot}$) of 8.09 $\pm$ 0.73 (standard deviation) and 8.07 $\pm$ 0.59 (standard deviation), respectively. Though there is large uncertainty in individual $M_{\rm BH,[O\,III]/nH\beta}$ measurements, the BM19 method provides a reasonable estimate of $M_{\rm BH}$ for a sample. {\it Right}: Eddington ratio ($L_{\rm bol}/L_{\rm Edd}$) distribution for our $M_{\rm BH,[O\,III]/nH\beta}$ sample and for the BASS obscured AGN sample. There is a much wider distribution of Eddington ratios for the BASS obscured AGN sample, indicating that the emission line ratio cuts we used remove AGN with extreme accretion rates, both at the low and high end.}
\end{figure*}

The distribution of $M_{\rm BH}$ and the Eddington ratio for both our $M_{\rm BH,[O\,III]/nH\beta}$ sample and the BASS obscured AGN sample are shown in Figure \ref{bass_comp}. The distribution of black hole masses calculated via the BM19 method is similar to that of the larger BASS obscured AGN sample calculated via $\sigma_{\rm *}$ with a median $\log$($M_{\rm BH}/M_{\rm \odot})$ of 8.09 $\pm$ 0.73 (standard deviation) for our sample and 8.07 $\pm$0.59 for the BASS obscured AGN sample. There is a wider range in values of Eddington ratios for the BASS obscured AGN sample, indicating that the emission line ratio cuts we employ to create our $M_{\rm BH,[O\,III]/nH\beta}$ sample removes AGN accreting at low and high Eddington rates. The consistency between the distribution of $M_{\rm BH}$ between our sample and the BASS obscured AGN sample indicates that though there is large uncertainty in the black hole mass for any individual AGN, the BM19 method provides a reasonable estimate of black hole masses for a population when other methods to measure $M_{\rm BH}$ are unavailable.

\section{Conclusions}
Using a sample of local AGN from the hard X-ray selected BASS survey \citep{koss2017,koss2022a} with intrinsic X-ray flux \citep{riccic2017}, optical spectroscopy \citep{oh2022,koss2022b}, and infrared spectroscopy \citep{lamperti2017,riccif2022,denbrok2022} measurements, we tested the proposed method from BM19 to measure black hole masses in Type 2 (obscured) AGN that have ionization ratios between 0.55 dex $<$ $\log$($L_{\rm [O\,III]}/L_{\rm nH\beta}$) $<$ 1.05 dex. This technique requires proxies for the FWHM of H$\alpha$ and optical accretion disk luminosity ($\lambda L{\rm 5100 \angstrom}$) in the black hole mass formula (Equation \ref{o3_hb_bhmass_formula}) since these parameters are not visible in obscured AGN. BM19 demonstrated that the ionization field hardness in the extended Narrow Line Region (parameterized by $L_{\rm [O\, III]}/L_{\rm nH\beta}$) correlates with the kinematics of gas in the Broad Line Region and can thus be used as a proxy of FWHM$_{\rm bH,\alpha}$ (Equation \ref{hafwhm_eq}).

Using a sample of 99 Type 2 AGN, we compared two methods for estimating $\lambda L{\rm 5100\angstrom}$, one based on the optical narrow emission lines \citep{netzer2009, runnoe2012} and the other based on the intrinsic 2-10 keV luminosity \citep{maiolino2007,lusso2010}. We find an average offset of 0.32$\pm$0.68 dex between these luminosity proxies. This scatter introduces an uncertainty of a factor of $\sim$2 when using the BM19 relationship. About half the uncertainty in black hole masses is due to this scatter. The other major contributor to black hole mass uncertainty results from the errors in the slope and normalization in the relationship to derive FWHM$_{bH\alpha}$ from the $L_{\rm [O\, III]}/L_{\rm nH\beta}$ ratio (Equation \ref{hafwhm_eq}).

Using Equations \ref{o3_hb_bhmass_formula}, \ref{hafwhm_eq}, and \ref{l5100_xray}, we calculated black hole masses ($M_{\rm BH,[OIII]/nH\beta}$) that we compared with virial mass measurements derived from the broad NIR Paschen lines \citep{kim2010,denbrok2022}. For the 14 (12) AGN that have reliable broad Pa$\alpha$ (Pa$\beta$) emission line measurements (Figure \ref{mbho3hb_mbhpa}), we found average offsets of  0.39$\pm$0.44 dex (0.48$\pm$0.51 dex). Though the offset is within the scatter, and the 95\% confidence interval on the linear fit overlaps the one-to-one relation, the black hole masses derived from the BM19 technique from this limited sample appear to be systematically higher than those calculated from the broad Paschen lines. There is tentative evidence that the Pa$\beta$-derived black hole masses may be biased low since $M_{\rm BH,Pa\beta}$ is systematically lower than both $M_{\rm BH,[OIII]/nH\beta}$ and black hole masses derived from the stellar velocity dispersion. More data are needed to test whether the apparent offset in $M_{\rm BH,[OIII]/nH\beta}$ is due to a small sample size.  The dispersion between $M_{\rm BH,[O\, III]/nH\beta}$ and  $M_{\rm BH,Pa\alpha}$ aligns (albeit on the high end) with both the intrinsic scatter seen in broad line single epoch spectrum virial mass calibrations \citep[e.g.,][]{vestergaard2006} and intercomparison of black hole masses from the broad line single epoch formulas \citep[e.g.,][]{trakhtenbrot2012,shen2012}.

When comparing $M_{\rm BH,[OIII]/nH\beta}$ with black hole masses derived from the stellar velocity dispersion \citep{kormendy2013,koss2022b} for a sample of 151 AGN, there is better overall agreement (mean offset of 0.08 dex) though with much wider scatter (0.74 dex). This scatter is a factor of about 2-3 higher than the observed scatter in $M_{\rm BH}$ - $\sigma_{\rm *}$ relationships \citep[e.g.][]{marsden2020} and larger than the typical uncertainty of 0.5 dex ascribed to the BM19 method based on their quoted scatter in $M_{\rm BH,[OIII]/nH\beta}$ - $\sigma_{\rm *}$.

We compare the black hole mass and Eddington ratio distribution for our  $M_{\rm BH,[OIII]/nH\beta}$ sample and a larger sample of obscured AGN from BASS, where $M_{\rm BH}$ was calculated from $\sigma_{\rm *}$. The distributions are similar with nearly identical median $M_{\rm BH}$ values. This result indicates that the BM19 method gives a reasonable estimate of $M_{\rm BH}$ on the population level even if individual measurements have large uncertainties.

A number of theoretical studies have compared $M_{\rm BH}$ to host galaxy properties to glean insight into black hole and galaxy co-evolution, and compared these predictions with observational results that include $M_{\rm BH,[OIII]/nH\beta}$  values from the sample of 10,000 SDSS Type 2 AGN reported in BM19 \citep{shankar2020,volenteri2020,habouzit2021,dubois2021,treibitsch2021,trinca2022,sassano2023,beckmann2023}. A handful of other papers use the methodology described in BM19 to estimate black hole masses for Type 2 AGN \citep{rey2021,ferre-mateu2021,vietri2022,siudek2023}. By testing this indirect method with a more direct measurement of black hole mass from the NIR Paschen lines, we find that the scatter is on the high end of that observed when comparing often-used single epoch spectroscopy broad line virial mass formulas. Our results also suggest that  $M_{\rm BH,[OIII]/nH\beta}$ may be biased high compared with $M_{\rm BH}$ from the broad Paschen lines, though the offset is within the scatter and more data would be needed to confirm this tentative result. We conclude that Equation \ref{hafwhm_eq} and Equation \ref{l5100_xray} (or Equations \ref{l5100_opt}, \ref{lbol}, \ref{ext_corr}, if the intrinsic 2-10 X-ray luminosity is unknown) can be used to {\it estimate} the black hole mass in obscured AGN when no other methods are feasible, though with the caveat that the uncertainty can be as high as 0.5 - 0.74 dex and the results may be biased high by a factor of $\sim 2- 3$.

We also caution that the correlation between $L_{\rm [O\,III]}/L_{\rm nH\beta}$ and FWHM$_{\rm bH,\alpha}$, which is the foundation for this BM19 $M_{\rm BH}$ estimate, has only been demonstrated for a limited parameter space: AGN with ionization ratios between 0.55 dex $<$ $\log$($L_{\rm [O\,III]}/L_{\rm nH\beta}$) $<$ 1.05 dex that are not hosted in low-metallicity galaxies \citep[see, e.g.,][]{hirschmann2019,dors2023}, and are below a redshift of $z < 1$ \citep[i.e., the $L_{\rm [O\,III]}/L_{\rm nH\beta}$ ratio for star-forming galaxies increases with redshift, where local relations no longer hold starting at $z \sim 1$;][]{kewley2013}. This technique can be useful for providing black hole mass estimates for AGN detected in X-ray surveys and for obscured AGN discovered in spectrocopic surveys that have measurements of [\ion{O}{3}], narrow H$\beta$, and narrow H$\alpha$ within this parameter space. A similar investigation into the correlation between $L_{\rm [O\,III]}/L_{\rm nH\beta}$ and FWHM$_{\rm bH,\alpha}$ can be done in the future for AGN at higher redshift to identify whether such a trend as that reported in BM19 for lower-redshift AGN is present, and, if so, what range of $L_{\rm [O\,III]}/L_{\rm nH\beta}$ would be appropriate as a proxy of BLR kinematics for estimating black hole masses in high-redshift obscured AGN.

\begin{acknowledgments}
We thank the referees for insightful feedback that improved the quality of this manuscript. B.T. acknowledges support from the European Research Council (ERC) under the European Union’s Horizon 2020 Research and Innovation program (grant agreement number 950533) and from the Israel Science Foundation (grant number 1849/19).
\end{acknowledgments}

%


\software{astropy \citep{astropy2013,astropy2018,astropy2022}
          }

\appendix
We provide a derivation for Equation \ref{l5100_xray} in the main text which we use to estimate $\lambda L_{\rm 5100 \angstrom}$ from the intrinsic 2-10 keV X-ray luminosity ($L_{\rm 2-10 keV}$). We begin with Equation 6 of \citet{lusso2010} which relates the monochromatic X-ray luminosity at 2 keV ($L_{\rm 2 keV}$) to the monochromatic UV luminosity at 2500 $\angstrom$ ($L_{\rm 2500 \angstrom}$):
\begin{equation}
\rm{\log} \left(\frac{L_{\rm 2keV}}{erg \, s^{-1} \, Hz^{-1}}\right) = 0.760 \times \rm{\log} \left(\frac{L_{\rm 2500}}{{\rm erg \, s^{-1} \, Hz^{-1}}} \right) + 3.508.
\end{equation}

This relation was calibrated using an X-ray selected sample of Type 1 AGN from the {\it XMM}-COSMOS survey \citep{hasinger2007,cappelluti2009}.

To calculate the monochromatic luminosity at 5100 $\angstrom$, we use the UV-optical spectral slope calculated from the mean spectrum of Type 1 AGN in SDSS that is reported in \citet{vandenberk}, $\alpha_{\nu}$ = -0.44, where $f_{\nu}\propto \lambda^{-(\alpha_{\nu} + 2)}$ :
\begin{equation}
  \begin{split}
  \frac{L_{\rm 2500}}{L_{\rm 5100}} = \left(\frac{2500}{5100}\right)^{-1(\alpha_{\nu} + 2)} = \left(\frac{2500}{5100}\right)^{-1.56} = 3.041
  \end{split}
\end{equation}

Substituting this relationship into the \citet{lusso2010} relation gives:
\begin{equation}\label{a1}
\rm{\log} \left(\frac{L_{\rm 2keV}}{erg \, s^{-1} \, Hz^{-1}}\right) = 0.760 \times \rm{\log} \left(\frac{L_{\rm 5100}}{{\rm erg \, s^{-1} \, Hz^{-1}}} \right) + 3.875
\end{equation}

We then convert from monochromatic luminosity to $\lambda L_{\rm 5100 \angstrom}$ (in units of erg s$^{-1}$) using:
\begin{equation}\label{a2}
  \left(\frac{\lambda L_{\rm 5100}}{{\rm erg \, s^{-1}}} \right) = \left(\frac{L_{\rm 5100}}{{\rm erg \, s^{-1} \, Hz^{-1}}} \right) \times \nu_{\rm5100} = \left(\frac{L_{\rm 5100}}{{\rm erg \, s^{-1} \, Hz^{-1}}} \right) \times 5.878 \times 10^{14} {\rm Hz}
\end{equation}

We convert the monochromatic 2 keV luminosity from units of erg s$^{-1}$ Hz$^{-1}$ to units of erg s$^{-1}$ keV$^{-1}$ using:
\begin{equation}\label{a3}
\left(\frac{L_{\rm 2keV}}{\rm erg \, s^{-1} \, keV^{-1}}\right) = \left(\frac{L_{\rm 2keV}}{\rm erg \, s^{-1} \, Hz^{-1}}\right) \times 2.418\times 10^{17} {\rm \frac{Hz}{keV}}
\end{equation}

To convert from monochromatic 2 keV luminosity to integrated 2-10 keV luminosity, we assume a standard X-ray spectral model for AGN where the X-ray spectral slope ($\Gamma$) is 1.8 which is the median $\Gamma$ for Swift BAT AGN \citep{riccic2017}.  Using \href{https://cxc.harvard.edu/toolkit/pimms.jsp}{PIMMS}, we find:
\begin{equation}\label{a4}
  L_{\rm 2-10 keV} = 3.80 \times L_{\rm 2 keV}
\end{equation}

Substituting Equations \ref{a2}, \ref{a3}, and \ref{a4} into \ref{a1}, we derive:
\begin{equation}
\rm{\log}\left(\frac{\lambda L_{\rm 5100}}{{\rm erg \, s^{-1}}} \right) = 1.316 \times  \rm{\log} \left(\frac{L_{\rm 2-10keV}}{\rm erg \, s^{-1}}\right)  - 13.966
\end{equation}

Finally, recasting this relation using normalized luminosities, we have:
\begin{equation}\label{l5100_appendix}
\rm{\log}\left(\frac{\lambda L_{\rm 5100}}{{\rm 10^{44} \, erg \, s^{-1}}} \right) =  1.316 \times \rm{\log} \left(\frac{L_{\rm 2-10keV}}{\rm 10^{43} \, erg \, s^{-1}}\right) -1.378
\end{equation}

We note that the estimate of $\lambda L_{\rm 5100 \angstrom}$ that is derived from the intrinsic X-ray luminosity has a weak dependence on the assumed spectral slope of the X-ray powerlaw emission. Within the range of typical AGN $\Gamma$ values of 1.7 \citep[e.g.,][]{maiolino2007} and 2.0 \citep[e.g.,][]{mainieri2007}, which also represent the range of the median fitted spectral slope for different subsets of BAT AGN separated by column density \citep{riccic2017} and AGN detected in other multi-wavelength surveys like {\it XMM}-XXL \citep{liu2016}, {\it Chandra} Deep Field South \citep{liu2017}, and Stripe 82X \citep{peca2023}, the normalization in Equation \ref{l5100_appendix} ranges from -1.426 ($\Gamma$ = 1.7) to -1.283 ($\Gamma$ = 2.0). Consequently, log($\lambda L_{\rm 5100 \angstrom}$) can vary by a factor of 0.14 dex which affects the estimated black hole mass by about 20\% (0.08 dex). 


\begin{thebibliography}

\bibitem[Abazajian et al.(2009)]{sdssdr7} Abazajian, K.~N., Adelman-McCarthy, J.~K., Ag{\"u}eros, M.~A., et al.\ 2009, \apjs, 182, 543. doi:10.1088/0067-0049/182/2/543

\bibitem[Alb{\'a}n \& Wylezalek(2023)]{alban2023} Alb{\'a}n, M. \& Wylezalek, D.\ 2023, \aap, 674, A85. doi:10.1051/0004-6361/202245437

\bibitem[Almeida et al.(2023)]{almeida2023} Almeida, A., Anderson, S.~F., Argudo-Fern{\'a}ndez, M., et al.\ 2023, \apjs, 267, 44. doi:10.3847/1538-4365/acda98

\bibitem[Astropy Collaboration et al.(2013)]{astropy2013} Astropy Collaboration, Robitaille, T.~P., Tollerud, E.~J., et al.\ 2013, \aap, 558, A33. doi:10.1051/0004-6361/201322068

\bibitem[Astropy Collaboration et al.(2018)]{astropy2018} Astropy Collaboration, Price-Whelan, A.~M., Sip{\H{o}}cz, B.~M., et al.\ 2018, \aj, 156, 123. doi:10.3847/1538-3881/aabc4f

\bibitem[Astropy Collaboration et al.(2022)]{astropy2022} Astropy Collaboration, Price-Whelan, A.~M., Lim, P.~L., et al.\ 2022, \apj, 935, 167. doi:10.3847/1538-4357/ac7c74

\bibitem[Baldwin et al.(1981)]{bpt} Baldwin, J.~A., Phillips, M.~M., \& Terlevich, R.\ 1981, \pasp, 93, 5. doi:10.1086/130766

\bibitem[Baron \& M{\'e}nard(2019)]{baron} Baron, D. \& M{\'e}nard, B.\ 2019, (BM19) \mnras, 487, 3404. doi:10.1093/mnras/stz1546

\bibitem[Barthelmy et al.(2005)]{barthelmy2005} Barthelmy, S.~D., Barbier, L.~M., Cummings, J.~R., et al.\ 2005, \ssr, 120, 143. doi:10.1007/s11214-005-5096-3

\bibitem[Bassani et al.(1999)]{bassani1999} Bassani, L., Dadina, M., Maiolino, R., et al.\ 1999, \apjs, 121, 473. doi:10.1086/313202

\bibitem[Beckmann et al.(2023)]{beckmann2023} Beckmann, R.~S., Dubois, Y., Volonteri, M., et al.\ 2023, \mnras, 523, 5610. doi:10.1093/mnras/stad1544

\bibitem[Bentz et al.(2006)]{bentz2006} Bentz, M.~C., Peterson, B.~M., Pogge, R.~W., et al.\ 2006, \apj, 644, 133. doi:10.1086/503537

\bibitem[Bentz et al.(2009)]{bentz2009} Bentz, M.~C., Peterson, B.~M., Netzer, H., et al.\ 2009, \apj, 697, 160. doi:10.1088/0004-637X/697/1/160

\bibitem[Bentz et al.(2013)]{bentz2013} Bentz, M.~C., Denney, K.~D., Grier, C.~J., et al.\ 2013, \apj, 767, 149. doi:10.1088/0004-637X/767/2/149

\bibitem[Braatz \& Gugliucci(2008)]{braatz2008} Braatz, J.~A. \& Gugliucci, N.~E.\ 2008, \apj, 678, 96. doi:10.1086/529538

\bibitem[Cappellari \& Emsellem(2004)]{cappellari2004} Cappellari, M. \& Emsellem, E.\ 2004, \pasp, 116, 138. doi:10.1086/381875

\bibitem[Cappelluti et al.(2009)]{cappelluti2009} Cappelluti, N., Brusa, M., Hasinger, G., et al.\ 2009, \aap, 497, 635. doi:10.1051/0004-6361/200810794
  
\bibitem[Cardelli et al.(1989)]{cardelli1989} Cardelli, J.~A., Clayton, G.~C., \& Mathis, J.~S.\ 1989, \apj, 345, 245. doi:10.1086/167900

\bibitem[Charlot \& Fall(2000)]{charlot2000} Charlot, S. \& Fall, S.~M.\ 2000, \apj, 539, 718. doi:10.1086/309250

\bibitem[den Brok et al.(2022)]{denbrok2022} den Brok, J.~S., Koss, M.~J., Trakhtenbrot, B., et al.\ 2022, \apjs, 261, 7. doi:10.3847/1538-4365/ac5b66

\bibitem[Dors et al.(2023)]{dors2023} Dors, O.~L., Cardaci, M.~V., Hagele, G.~F., et al.\ 2023, arXiv:2311.14026. doi:10.48550/arXiv.2311.14026

\bibitem[Dubois et al.(2021)]{dubois2021} Dubois, Y., Beckmann, R., Bournaud, F., et al.\ 2021, \aap, 651, A109. doi:10.1051/0004-6361/202039429

\bibitem[Ferrarese \& Merritt(2000)]{ferrarese2000} Ferrarese, L. \& Merritt, D.\ 2000, \apjl, 539, L9. doi:10.1086/312838

\bibitem[Ferrarese et al.(2001)]{ferrarese2001} Ferrarese, L., Pogge, R.~W., Peterson, B.~M., et al.\ 2001, \apjl, 555, L79. doi:10.1086/322528

\bibitem[Ferr{\'e}-Mateu et al.(2021)]{ferre-mateu2021} Ferr{\'e}-Mateu, A., Mezcua, M., \& Barrows, R.~S.\ 2021, \mnras, 506, 4702. doi:10.1093/mnras/stab1915

\bibitem[Gebhardt et al.(2000)]{gebhardt2000} Gebhardt, K., Bender, R., Bower, G., et al.\ 2000, \apjl, 539, L13. doi:10.1086/312840

\bibitem[Gehrels et al.(2004)]{gehrels2004} Gehrels, N., Chincarini, G., Giommi, P., et al.\ 2004, \apj, 611, 1005. doi:10.1086/422091

\bibitem[Genzel et al.(2000)]{genzel2000} Genzel, R., Pichon, C., Eckart, A., et al.\ 2000, \mnras, 317, 348. doi:10.1046/j.1365-8711.2000.03582.x

\bibitem[Genzel et al.(2010)]{genzel2010} Genzel, R., Eisenhauer, F., \& Gillessen, S.\ 2010, Reviews of Modern Physics, 82, 3121. doi:10.1103/RevModPhys.82.3121

\bibitem[Ghez et al.(1998)]{ghez1998} Ghez, A.~M., Klein, B.~L., Morris, M., et al.\ 1998, \apj, 509, 678. doi:10.1086/306528

\bibitem[Ghez et al.(2008)]{ghez2008} Ghez, A.~M., Salim, S., Weinberg, N.~N., et al.\ 2008, \apj, 689, 1044. doi:10.1086/592738


\bibitem[Greene \& Ho(2005)]{greene2005} Greene, J.~E. \& Ho, L.~C.\ 2005, \apj, 630, 122. doi:10.1086/431897

\bibitem[Greene et al.(2010)]{greene2010} Greene, J.~E., Peng, C.~Y., \& Ludwig, R.~R.\ 2010, \apj, 709, 937. doi:10.1088/0004-637X/709/2/937

\bibitem[Greene et al.(2016)]{greene2016} Greene, J.~E., Seth, A., Kim, M., et al.\ 2016, \apjl, 826, L32. doi:10.3847/2041-8205/826/2/L32
  
\bibitem[Greenhill et al.(2003)]{greenhill2003} Greenhill, L.~J., Kondratko, P.~T., Lovell, J.~E.~J., et al.\ 2003, \apjl, 582, L11. doi:10.1086/367602

\bibitem[Grier et al.(2017)]{grier2017} Grier, C.~J., Trump, J.~R., Shen, Y., et al.\ 2017, \apj, 851, 21. doi:10.3847/1538-4357/aa98dc

\bibitem[Grier et al.(2019)]{grier2019} Grier, C.~J., Shen, Y., Horne, K., et al.\ 2019, \apj, 887, 38. doi:10.3847/1538-4357/ab4ea5


\bibitem[G{\"u}ltekin et al.(2009)]{gultekin2009} G{\"u}ltekin, K., Richstone, D.~O., Gebhardt, K., et al.\ 2009, \apj, 698, 198. doi:10.1088/0004-637X/698/1/198

\bibitem[Habouzit et al.(2021)]{habouzit2021} Habouzit, M., Li, Y., Somerville, R.~S., et al.\ 2021, \mnras, 503, 1940. doi:10.1093/mnras/stab496

\bibitem[Hasinger et al.(2007)]{hasinger2007} Hasinger, G., Cappelluti, N., Brunner, H., et al.\ 2007, \apjs, 172, 29. doi:10.1086/516576

\bibitem[Hickox \& Alexander(2018)]{hickox2018} Hickox, R.~C. \& Alexander, D.~M.\ 2018, \araa, 56, 625. doi:10.1146/annurev-astro-081817-051803
  
\bibitem[Hirschmann et al.(2019)]{hirschmann2019} Hirschmann, M., Charlot, S., Feltre, A., et al.\ 2019, \mnras, 487, 333. doi:10.1093/mnras/stz1256

\bibitem[Homayouni et al.(2020)]{homayouni2020} Homayouni, Y., Trump, J.~R., Grier, C.~J., et al.\ 2020, \apj, 901, 55. doi:10.3847/1538-4357/ababa9
  
\bibitem[Kaspi et al.(2005)]{kaspi2005} Kaspi, S., Maoz, D., Netzer, H., et al.\ 2005, \apj, 629, 61. doi:10.1086/431275

\bibitem[Kewley et al.(2001)]{kewley2001} Kewley, L.~J., Dopita, M.~A., Sutherland, R.~S., et al.\ 2001, \apj, 556, 121. doi:10.1086/321545
  
\bibitem[Kewley et al.(2013)]{kewley2013} Kewley, L.~J., Maier, C., Yabe, K., et al.\ 2013, \apjl, 774, L10. doi:10.1088/2041-8205/774/1/L10
  
\bibitem[Kim et al.(2010)]{kim2010} Kim, D., Im, M., \& Kim, M.\ 2010, \apj, 724, 386. doi:10.1088/0004-637X/724/1/386

\bibitem[Kormendy \& Richstone(1995)]{kormendy1995} Kormendy, J. \& Richstone, D.\ 1995, \araa, 33, 581. doi:10.1146/annurev.aa.33.090195.003053

\bibitem[Kormendy \& Ho(2013)]{kormendy2013} Kormendy, J. \& Ho, L.~C.\ 2013, \araa, 51, 511. doi:10.1146/annurev-astro-082708-101811

\bibitem[Koss et al.(2017)]{koss2017} Koss, M., Trakhtenbrot, B., Ricci, C., et al.\ 2017, \apj, 850, 74. doi:10.3847/1538-4357/aa8ec9

\bibitem[Koss et al.(2022a)]{koss2022a} Koss, M.~J., Trakhtenbrot, B., Ricci, C., et al.\ 2022, \apjs, 261, 1. doi:10.3847/1538-4365/ac6c8f

\bibitem[Koss et al.(2022b)]{koss2022b} Koss, M.~J., Trakhtenbrot, B., Ricci, C., et al.\ 2022, \apjs, 261, 6. doi:10.3847/1538-4365/ac650b

\bibitem[Koss et al.(2022c)]{koss2022c} Koss, M.~J., Ricci, C., Trakhtenbrot, B., et al.\ 2022, \apjs, 261, 2. doi:10.3847/1538-4365/ac6c05

\bibitem[LaMassa et al.(2009)]{lamassa2009} LaMassa, S.~M., Heckman, T.~M., Ptak, A., et al.\ 2009, \apj, 705, 568. doi:10.1088/0004-637X/705/1/568

\bibitem[LaMassa et al.(2010)]{lamassa2010} LaMassa, S.~M., Heckman, T.~M., Ptak, A., et al.\ 2010, \apj, 720, 786. doi:10.1088/0004-637X/720/1/786
  
\bibitem[LaMassa et al.(2023)]{lamassa2023} LaMassa, S.~M., Yaqoob, T., Tzanavaris, P., et al.\ 2023, \apj, 944, 152. doi:10.3847/1538-4357/acb3bb

 \bibitem[Lamperti et al.(2017)]{lamperti2017} Lamperti, I., Koss, M., Trakhtenbrot, B., et al.\ 2017, \mnras, 467, 540. doi:10.1093/mnras/stx055

\bibitem[Landt et al.(2011a)]{landt2011a} Landt, H., Bentz, M.~C., Peterson, B.~M., et al.\ 2011, \mnras, 413, L106. doi:10.1111/j.1745-3933.2011.01047.x

\bibitem[Landt et al.(2011b)]{landt2011b} Landt, H., Elvis, M., Ward, M.~J., et al.\ 2011, \mnras, 414, 218. doi:10.1111/j.1365-2966.2011.18383.x

\bibitem[Landt et al.(2013)]{landt2013} Landt, H., Ward, M.~J., Peterson, B.~M., et al.\ 2013, \mnras, 432, 113. doi:10.1093/mnras/stt421

\bibitem[Liu et al.(2016)]{liu2016} Liu, Z., Merloni, A., Georgakakis, A., et al.\ 2016, \mnras, 459, 1602. doi:10.1093/mnras/stw753

\bibitem[Liu et al.(2017)]{liu2017} Liu, T., Tozzi, P., Wang, J.-X., et al.\ 2017, \apjs, 232, 8. doi:10.3847/1538-4365/aa7847

\bibitem[Lusso et al.(2010)]{lusso2010} Lusso, E., Comastri, A., Vignali, C., et al.\ 2010, \aap, 512, A34. doi:10.1051/0004-6361/200913298

\bibitem[Mainieri et al.(2007)]{mainieri2007} Mainieri, V., Hasinger, G., Cappelluti, N., et al.\ 2007, \apjs, 172, 368. doi:10.1086/516573
  
\bibitem[Maiolino et al.(2001)]{maiolino2001} Maiolino, R., Marconi, A., Salvati, M., et al.\ 2001, \aap, 365, 28. doi:10.1051/0004-6361:20000177

\bibitem[Magorrian et al.(1998)]{magorrian1998} Magorrian, J., Tremaine, S., Richstone, D., et al.\ 1998, \aj, 115, 2285. doi:10.1086/300353

\bibitem[Maiolino et al.(2007)]{maiolino2007} Maiolino, R., Shemmer, O., Imanishi, M., et al.\ 2007, \aap, 468, 979. doi:10.1051/0004-6361:20077252

\bibitem[Marsden et al.(2020)]{marsden2020} Marsden, C., Shankar, F., Ginolfi, M., et al.\ 2020, Frontiers in Physics, 8, 61. doi:10.3389/fphy.2020.00061

\bibitem[Mej{\'\i}a-Restrepo et al.(2022)]{mejia-restrepo2022} Mej{\'\i}a-Restrepo, J.~E., Trakhtenbrot, B., Koss, M.~J., et al.\ 2022, \apjs, 261, 5. doi:10.3847/1538-4365/ac6602

\bibitem[Misner et al.(1973)]{misner} Misner, C.~W., Thorne, K.~S., \& Wheeler, J.~A.\ 1973, Gravitation.  By C.W. Misner, K.S. Thorne and J.A. Wheeler. ISBN 0-7167-0334-3, ISBN 0-7167-0344-0 (pbk). San Francisco: W.H. Freeman and Company, 1973

\bibitem[Miyoshi et al.(1995)]{miyoshi1995} Miyoshi, M., Moran, J., Herrnstein, J., et al.\ 1995, \nat, 373, 127. doi:10.1038/373127a0

\bibitem[Moran et al.(1999)]{moran1999} Moran, J.~M., Greenhill, L.~J., \& Herrnstein, J.~R.\ 1999, Journal of Astrophysics and Astronomy, 20, 165. doi:10.1007/BF02702350

\bibitem[Netzer(2009)]{netzer2009} Netzer, H.\ 2009, \mnras, 399, 1907. doi:10.1111/j.1365-2966.2009.15434.x

\bibitem[Oh et al.(2022)]{oh2022} Oh, K., Koss, M.~J., Ueda, Y., et al.\ 2022, \apjs, 261, 4. doi:10.3847/1538-4365/ac5b68

\bibitem[Osterbrock(1993)]{osterbrock1993} Osterbrock, D.~E.\ 1993, \apj, 404, 551. doi:10.1086/172307


\bibitem[Osterbrock \& Ferland(2006)]{osterbrock2006} Osterbrock, D.~E. \& Ferland, G.~J.\ 2006, Astrophysics of gaseous nebulae and active galactic nuclei, 2nd. ed. by D.E. Osterbrock and G.J. Ferland. Sausalito, CA: University Science Books, 2006

\bibitem[Peca et al.(2023)]{peca2023} Peca, A., Cappelluti, N., Urry, C.~M., et al.\ 2023, \apj, 943, 162. doi:10.3847/1538-4357/acac28

\bibitem[Peterson(2014)]{peterson2014} Peterson, B.~M.\ 2014, \ssr, 183, 253. doi:10.1007/s11214-013-9987-4

\bibitem[Planck Collaboration et al.(2016)]{planck2015} Planck Collaboration, Ade, P.~A.~R., Aghanim, N., et al.\ 2016, \aap, 594, A13. doi:10.1051/0004-6361/201525830

\bibitem[Ramos Almeida \& Ricci(2017)]{ramosalmeida2017} Ramos Almeida, C. \& Ricci, C.\ 2017, Nature Astronomy, 1, 679. doi:10.1038/s41550-017-0232-z

\bibitem[Reines \& Volonteri(2015)]{reines2015} Reines, A.~E. \& Volonteri, M.\ 2015, \apj, 813, 82. doi:10.1088/0004-637X/813/2/82

\bibitem[Rey et al.(2021)]{rey2021} Rey, S.-C., Oh, K., \& Kim, S.\ 2021, \apjl, 917, L9. doi:10.3847/2041-8213/ac15f6

\bibitem[Ricci et al.(2017)]{riccic2017} Ricci, C., Trakhtenbrot, B., Koss, M.~J., et al.\ 2017, \apjs, 233, 17. doi:10.3847/1538-4365/aa96ad

\bibitem[Ricci et al.(2022)]{riccif2022} Ricci, F., Treister, E., Bauer, F.~E., et al.\ 2022, \apjs, 261, 8. doi:10.3847/1538-4365/ac5b67

\bibitem[Ricci et al.(2017)]{riccif2017} Ricci, F., La Franca, F., Onori, F., et al.\ 2017, \aap, 598, A51. doi:10.1051/0004-6361/201629380

\bibitem[Runnoe et al.(2012)]{runnoe2012} Runnoe, J.~C., Brotherton, M.~S., \& Shang, Z.\ 2012, \mnras, 422, 478. doi:10.1111/j.1365-2966.2012.20620.x

\bibitem[Sassano et al.(2023)]{sassano2023} Sassano, F., Capelo, P.~R., Mayer, L., et al.\ 2023, \mnras, 519, 1837. doi:10.1093/mnras/stac3608

\bibitem[Shankar et al.(2020)]{shankar2020} Shankar, F., Allevato, V., Bernardi, M., et al.\ 2020, Nature Astronomy, 4, 282. doi:10.1038/s41550-019-0949-y

\bibitem[Shen et al.(2011)]{shen2011} Shen, Y., Richards, G.~T., Strauss, M.~A., et al.\ 2011, \apjs, 194, 45. doi:10.1088/0067-0049/194/2/45

\bibitem[Shen \& Liu(2012)]{shen2012} Shen, Y. \& Liu, X.\ 2012, \apj, 753, 125. doi:10.1088/0004-637X/753/2/125

\bibitem[Shen et al.(2015)]{shen2015} Shen, Y., Brandt, W.~N., Dawson, K.~S., et al.\ 2015, \apjs, 216, 4. doi:10.1088/0067-0049/216/1/4

\bibitem[Shen et al.(2019)]{shen2019} Shen, Y., Hall, P.~B., Horne, K., et al.\ 2019, \apjs, 241, 34. doi:10.3847/1538-4365/ab074f

\bibitem[Shen et al.(2024)]{shen2024} Shen, Y., Grier, C.~J., Horne, K., et al.\ 2024, \apjs, 272, 26. doi:10.3847/1538-4365/ad3936

\bibitem[Siudek et al.(2023)]{siudek2023} Siudek, M., Mezcua, M., \& Krywult, J.\ 2023, \mnras, 518, 724. doi:10.1093/mnras/stac3092
  
\bibitem[Steffen et al.(2006)]{steffen2006} Steffen, A.~T., Strateva, I., Brandt, W.~N., et al.\ 2006, \aj, 131, 2826. doi:10.1086/503627

\bibitem[Tananbaum et al.(1979)]{tananbaum1979} Tananbaum, H., Avni, Y., Branduardi, G., et al.\ 1979, \apjl, 234, L9. doi:10.1086/183100

\bibitem[Trakhtenbrot \& Netzer(2012)]{trakhtenbrot2012} Trakhtenbrot, B. \& Netzer, H.\ 2012, \mnras, 427, 3081. doi:10.1111/j.1365-2966.2012.22056.x

\bibitem[Trebitsch et al.(2021)]{treibitsch2021} Trebitsch, M., Dubois, Y., Volonteri, M., et al.\ 2021, \aap, 653, A154. doi:10.1051/0004-6361/202037698

\bibitem[Trinca et al.(2022)]{trinca2022} Trinca, A., Schneider, R., Valiante, R., et al.\ 2022, \mnras, 511, 616. doi:10.1093/mnras/stac062
  
\bibitem[Vanden Berk et al.(2001)]{vandenberk} Vanden Berk, D.~E., Richards, G.~T., Bauer, A., et al.\ 2001, \aj, 122, 549. doi:10.1086/321167

\bibitem[Vasudevan \& Fabian(2009)]{vasudevan2009} Vasudevan, R.~V. \& Fabian, A.~C.\ 2009, \mnras, 392, 1124. doi:10.1111/j.1365-2966.2008.14108.x

\bibitem[Vestergaard \& Peterson(2006)]{vestergaard2006} Vestergaard, M. \& Peterson, B.~M.\ 2006, \apj, 641, 689. doi:10.1086/500572

\bibitem[Vietri et al.(2022)]{vietri2022} Vietri, G., Garilli, B., Polletta, M., et al.\ 2022, \aap, 659, A129. doi:10.1051/0004-6361/202141072

\bibitem[Volonteri et al.(2020)]{volenteri2020} Volonteri, M., Pfister, H., Beckmann, R.~S., et al.\ 2020, \mnras, 498, 2219. doi:10.1093/mnras/staa2384

\bibitem[Woo et al.(2015)]{woo2015} Woo, J.-H., Yoon, Y., Park, S., et al.\ 2015, \apj, 801, 38. doi:10.1088/0004-637X/801/1/38

\bibitem[York et al.(2000)]{york2000} York, D.~G., Adelman, J., Anderson, J.~E., et al.\ 2000, \aj, 120, 1579. doi:10.1086/301513

\bibitem[Zhu et al.(2011)]{zhu2011} Zhu, G., Zaw, I., Blanton, M.~R., et al.\ 2011, \apj, 742, 73. doi:10.1088/0004-637X/742/2/73

  
\end{thebibliography}
\end{document}